\documentclass[12pt,letterpaper]{article}

\usepackage{amsmath}
\usepackage{amssymb}
\usepackage{epsfig}
\usepackage{graphicx}
\usepackage{cite}
\newcommand{\capdef}{}
\newcommand{\mycaption}[2][\capdef]{\renewcommand{\capdef}{#2}%
       \caption[#1]{{\footnotesize #2}}}
\makeatletter
\renewcommand{\fnum@table}{\textbf{\tablename~\thetable}}
\renewcommand{\fnum@figure}{\textbf{\figurename~\thefigure}}
\makeatother

\newcounter{myenumi}

\renewcommand{\themyenumi}{\roman{myenumi}}
{\end{list}}

\newlength{\myem}
\newcounter{mysubequation}[equation]

\makeatletter
\renewcommand{\section}{\@startsection{section}{1}{0em}{-\baselineskip}%
{\baselineskip}{\normalfont\large\bfseries}}
\renewcommand{\subsection}%
{\@startsection{subsection}{2}{0em}{-0.7\baselineskip}%
{0.7\baselineskip}{\normalfont\bfseries}}
\makeatother

\newcommand{\ie}{{\it i.e.}}

\newcommand{\eg}{{\it e.g.}}

\newcommand{\cf}{{\it cf.}}
\newcommand{\etc}{{\it etc.}}
\newcommand{\eq}{Eq.}

\newcommand{\fig}{Figure}

\newcommand{\Ref}{Ref.}
\newcommand{\Refs}{Refs.}
\newcommand{\Sec}{Section}

\newcommand{\Tab}{Table}

\newcommand{\JHFSK}{\mbox{\sf T2K}}
\newcommand{\JHFHK}{\mbox{\sf T2HK}}
\newcommand{\NuFactII}{\mbox{\sf NuFact-II}}

\newcommand{\NUMI}{\mbox{\sf NO$\nu$A}}
\newcommand{\MINOS}{\mbox{\sf MINOS}}
\newcommand{\ICARUS}{\mbox{\sf ICARUS}}
\newcommand{\OPERA}{\mbox{\sf OPERA}}

\newcommand{\ReactorII}{\mbox{\sf Reactor-II}}
\newcommand{\TEN}{\JHFSK +\NUMI +\ReactorII}

\newcommand{\stheta}{\sin^22\theta_{13}}
\newcommand{\deltacp}{\delta_\mathrm{CP}}
\newcommand{\ldm}{\Delta m_{31}^2}
\newcommand{\sdm}{\Delta m_{21}^2}
\newcommand{\equ}[1]{\eq~(\ref{equ:#1})}
\newcommand{\figu}[1]{\fig~\ref{fig:#1}}

\newcommand{\bi}{\begin{itemize}}
\newcommand{\ei}{\end{itemize}}

\newcommand{\cps}{%
Barger:1980jm,%
Arafune:1996bt,%
Tanimoto:1996by,%
Tanimoto:1996ky,%
Arafune:1997hd,%
Bilenky:1997dd,%
Koike:1997dh,%
Minakata:1997td,%
Minakata:1998bf,%
Tanimoto:1998sn,%
DeRujula:1998hd,%
Barger:1999fs,%
Dick:1999ed,%
Donini:1999jc,%
Freund:1999gy,%
Koike:1999hf,%
Romanino:1999zq,%
Sato:1999wt,%
Cervera:2000kp,%
Minakata:2000fe,%
Minakata:2000ee,%
Koike:2000jf,%
Barger:2001yr,%
Bueno:2001jd,%
Burguet-Castell:2001ez,%
Koike:2001kv,%
Minakata:2001rj,%
Pinney:2001xw,%
Aoki:2002ae,%
Barger:2002xk,%
Burguet-Castell:2002qx,%
Donini:2002rm,%
Huber:2002mx,%
Huber:2002rs,%
Minakata:2002qi,%
Minakata:2002qe,%
Ota:2002fu,%
Whisnant:2002fx,%
Zucchelli:2002sa,%
Autiero:2003fu,%
Burguet-Castell:2003vv,%
Brahmachari:2003bk,%
Diwan:2003bp,%
Donini:2003vz,%
Huber:2003ak,%
Mezzetto:2003ub,%
Migliozzi:2003pw,%
Minakata:2003wq,%
Ohlsson:2003ip,%
Shan:2003vh,%
Winter:2003ye,%
Blom:2004bk,%
Huber:2004ug,%
Mena:2004sa,%
Minakata:2004vz%
}

\newcommand{\HCPI}{HCPI}

\begin{document}
%%%%%%%%%%%%%%%%%%%%%%%%%%%%%%%%%%%%%%%%%%%%%%%%%%%%%%%%%%%%%%%%%%%%%
%%%%                     Title-page                              %%%%
%%%%%%%%%%%%%%%%%%%%%%%%%%%%%%%%%%%%%%%%%%%%%%%%%%%%%%%%%%%%%%%%%%%%%

\begin{titlepage}

% the footnote symbols are only redefined for the title page !
\renewcommand{\thefootnote}{\alph{footnote}}

\vspace*{-3.cm}
\begin{flushright}
TUM-HEP-571/04\\
MADPH-04-1411
%hep-ph/
\end{flushright}

\vspace*{0.5cm}

\renewcommand{\thefootnote}{\fnsymbol{footnote}}
\setcounter{footnote}{-1}

%-ML: Waere es nicht besser measurements durch determinations zu ersetzen?

{\begin{center}
{\Large\bf From parameter space constraints to the precision determination of the leptonic Dirac CP phase}
\end{center}}
\renewcommand{\thefootnote}{\alph{footnote}}

\vspace*{.8cm}
%\vspace*{.3cm}
{\begin{center} {\large{\sc
                P.~Huber\footnote[1]{\makebox[1.cm]{Email:}
                phuber@physics.wisc.edu},~
                M.~Lindner\footnote[2]{\makebox[1.cm]{Email:}
                lindner@ph.tum.de},~
                W.~Winter\footnote[3]{\makebox[1.cm]{Email:}
                winter@ias.edu}
                }}
\end{center}}
\vspace*{0cm}
{\it
\begin{center}

\footnotemark[1]%${}^,$\footnotemark[2]%
       Department of Physics, University of Wisconsin, \\
       1150 University Avenue, Madison, WI 53706, USA

\vspace*{1mm}

\footnotemark[2]%
       Physik--Department, Technische Universit\"at M\"unchen, \\
       James--Franck--Strasse, 85748 Garching, Germany

\vspace*{1mm}

\footnotemark[3]%
       School of Natural Sciences, Institute for Advanced Study, \\
       Einstein Drive, Princeton, NJ 08540, USA

\vspace*{1cm}

\today
\end{center}}

\vspace*{0.3cm}

\begin{abstract}
We discuss the precision determination of the leptonic
Dirac CP phase $\deltacp$ in neutrino oscillation experiments, where we apply the
concept of  ``CP coverage''. We demonstrate that this approach
carries more information than a conventional CP violation measurement,
since it also describes the exclusion of parameter regions.
This will be very useful for next-generation long baseline experiments
where for sizable $\sin^2 2 \theta_{13}$ first constraints on
$\deltacp$ can be obtained. As the most sophisticated experimental
setup, we analyze neutrino factories, where we illustrate the major
difficulties in their analysis. In addition, we compare their
potential to the one of superbeam upgrades and next-generation
experiments, which also includes a discussion of synergy effects.
We find a strong dependence on the yet unknown true values of
$\sin^2 2 \theta_{13}$ and $\deltacp$, as well as a strong, non-Gaussian
dependence on the confidence level. A systematic understanding of the
complicated parameter dependence will be given. In addition, it is
shown that comparisons of experiments and synergy discussions do in
general not allow for an unbiased judgment if they are only performed
at selected points in parameter space. Therefore, we present our results
in dependence of the yet unknown true values of $\sin^2 2 \theta_{13}$
and $\deltacp$. Finally we show that for $\deltacp$ precision measurements
there exist simple strategies including superbeams, reactor experiments,
superbeam upgrades, and neutrino factories, where the crucial
discriminator is $\sin^2 2 \theta_{13} \sim 10^{-2}$.
\end{abstract}

\vspace*{.5cm}

\end{titlepage}

\newpage

\renewcommand{\thefootnote}{\arabic{footnote}}
\setcounter{footnote}{0}

%%%%%%%%%%%%%%%%%%%%%%%%%%%%%%%%%%%%%%%%%%%%%%%%%%%%%%%%%%%%%%%%%%%%%
%                     Introduction                                  %
%%%%%%%%%%%%%%%%%%%%%%%%%%%%%%%%%%%%%%%%%%%%%%%%%%%%%%%%%%%%%%%%%%%%%

\section{Introduction}

After the measurements of the leading atmospheric and solar oscillation
parameters (see, \eg\ \Ref~\cite{Maltoni:2004ei} and references therein),
the most important task for the next generation neutrino oscillation
experiments will be the search for a finite value of $\theta_{13}$.
Future reactor experiments~\cite{Anderson:2004pk}, conventional beam
experiments~\cite{Ables:1995wq,Duchesneau:2002yq,Aprili:2002wx}, or superbeams~\cite{Itow:2001ee,Ayres:2002nm} could
establish $\stheta>0$ for $\stheta \gtrsim 10^{-2}$. For smaller
values of $\theta_{13}$, superbeam upgrades (such as \JHFHK ~\cite{Itow:2001ee}) and $\beta$-beams~\cite{Zucchelli:2002sa,Burguet-Castell:2003vv}, or neutrino factories~\cite{Geer:1998iz,Apollonio:2002en,Albright:2004iw} continue
the hunt.  Once $\stheta >0$ is established, it is possible to address
the mass hierarchy and the value of $\deltacp$. In this study we discuss
precision measurements of the leptonic Dirac CP phase as the most
challenging task for neutrino oscillation physics within the framework
of three-flavor neutrino oscillations.

The determination of the leptonic CP phase $\deltacp$ has so far been
discussed in different ways~\cite{\cps}. 
Some of them are CP violation measurements, which address the question
if a CP violating value of $\deltacp$ can be distinguished from the CP
conserving values $0$ and $\pi$. The sensitivity to maximal CP violation
$\deltacp=\pi/2$ or $\deltacp=3/2 \, \pi$ has also been extensively
studied. Another class are precision measurements of $\deltacp$, which
have been investigated \eg\ in the context of superbeam upgrades and
neutrino factories.
Many results are given only at some discrete points in parameter
space and from this it is not possible to judge the overall performance
of the considered setup. A discussion of the changes in the precision
as function of the relevant parameter values is missing in most cases.
The aim of this study is therefore to investigate the reason for
differences in previous studies and to a find a systematic way for
comparing and classifying the CP sensitivities. We will demonstrate that
these differences are indeed consistent, and we will illustrate
how one can understand the underlying parameter dependence. In particular,
we will find that the topology of the neutrino factory factor parameter
space for small values of $\stheta$ is rather complicated.

From the theoretical point of view any value of $\deltacp$ could be
realized by nature, \ie\ $\deltacp \in [0,2 \pi[$.
This means, for example, that sensitivity to maximal CP violation
($\deltacp = \pi/2$ or $3/2 \, \pi$) does very likely not correspond
to the real world. In particular, the era of superbeams and reactor
experiments within the next ten years will not be able to measure CP
violation even under very optimistic assumptions~\cite{Huber:2004ug}.
Nevertheless, superbeams could exclude some values of $\deltacp$, and
thus could restrict the possible parameter space. On the other end, at
the high precision frontier, the investigation of CP violation might
be too restrictive, since these future experiments could not only
establish CP violation, but also constrain the parameter space for
$\deltacp$ even further. Therefore, all available information on
$\deltacp$ should be used, which then can be used for the optimization
of future experiments, as input for neutrino mass models, or as
motivation to continue hunting for leptonic CP violation.

One of the major difficulties in the analysis of future neutrino oscillation
experiments is the huge number of parameters: In general, one has six simulated
parameters (within their currently allowed ranges) and six fitted parameters,
\ie, a 12-dimensional parameter space. A performance indicator which
condenses the information is therefore required in order to show the results
as a function of the most relevant impact parameters. Such a performance
indicator can, furthermore, be used for the comparison of experiments, for risk
minimization with respect to the yet unknown parameter values, for the
optimization of experiments, or for the discussion of synergy effects. In
this study, we use the ``CP coverage''~\cite{Huber:2002mx} as
performance indicator for CP precision measurements.
CP coverage is defined as the combined range of all fit values which fit
the chosen simulated value of $\deltacp$.\footnote{In this study, we
use always degrees for the CP coverage (fit values), whereas we use radians for
simulated/true values of $\deltacp$.}
Thus, a very small CP coverage corresponds to a good precision of $\deltacp$,
whereas a CP coverage of $360^\circ$ corresponds to no information on
$\deltacp$. Note that this definition includes the case of disjoint regions (degeneracies)
irrespectively of the precision within each individual region. For example, a
value of $300^\circ$ means
that in total $60^\circ$ of the possible  parameter range of $360^\circ$ can
be excluded. Therefore, the CP coverage is a useful performance indicator
which interpolates between exclusion measurements and high precision
measurements of the leptonic Dirac CP phase.
Note that the CP coverage should not be confused with the often
used ``CP fraction'' (for example, in \Ref~\cite{Ambats:2004js}).
CP coverage refers to a range of fitted values of $\deltacp$, whereas
CP fraction refers to a range of simulated/true values.

This study is organized as follows: In \Sec~\ref{sec:framework}, we give a
short introduction to three-flavor neutrino oscillations and the appearance
channel in future long-baseline experiments. This section can be skipped by a
reader familiar with the subject. Then we describe in \Sec~\ref{sec:analysis}
the analysis methods, where we only quickly repeat the general techniques
described in earlier works, and rather extensively illustrate the problems
connected with the analysis of future high precision instruments. This section
is kept on a rather illustrative level, but is mandatory to understand the
more technical details in \Sec~\ref{sec:results1}. The quite
general first part of \Sec~\ref{sec:results1} introduces performance
indicators to differentiate experiment classes and to investigate synergies
and shows the consequences for the physics potential of different experiments.
The more technical second part of this section investigates the
specific characteristics of superbeam and neutrino factory parameter space.
The latter is not a prerequisite to understand \Sec~\ref{sec:results2}, which
 discusses possible synergies among experiment types for future $\deltacp$
precision measurements. In particular, it is a major objective of this
section to investigate the complete parameter space in a systematized manner.
Finally, we summarize our results in \Sec~\ref{sec:summary}.

\section{Neutrino oscillation framework}
\label{sec:framework}

For long-baseline beam experiments, the electron or muon neutrino
appearance probability $P_{\mathrm{app}}$ in matter carries the main
information for CP effects. It can be expanded in the small hierarchy
parameter $\alpha \equiv \Delta m_{21}^2/\Delta m_{31}^2$ and the
small $\sin 2 \theta_{13}$ up to the second order
as~\cite{Cervera:2000kp,Freund:2001pn,Akhmedov:2004ny}:
\begin{eqnarray}
P_{\mathrm{app}} & \simeq & \sin^2 2\theta_{13} \, \sin^2 \theta_{23}
\frac{\sin^2[(1- \hat{A}){\Delta}]}{(1-\hat{A})^2}
\nonumber \\
&\pm&   \alpha  \sin 2\theta_{13} \, \sin 2\theta_{12} \, \sin 2\theta_{23} 
\, \sin \delta_{\mathrm{CP}}
\sin({\Delta})  \frac{\sin(\hat{A}{\Delta})}{\hat{A}}  
\frac{\sin[(1-\hat{A}){\Delta}]}{(1-\hat{A})}
\nonumber  \\
&+&   \alpha  \sin 2\theta_{13} \,   \sin 2\theta_{12} \, \sin 2\theta_{23} \,
 \cos \delta_{\mathrm{CP}} \cos({\Delta})  
\frac{\sin(\hat{A}{\Delta})}{\hat{A}}  
\frac{\sin[(1-\hat{A}){\Delta}]} {(1-\hat{A})}
 \nonumber  \\
&+&  \alpha^2 \, \cos^2 \theta_{23}  \sin^2 2\theta_{12}
\frac{\sin^2(\hat{A}{\Delta})}{\hat{A}^2}.
\label{equ:PROBMATTER}
\end{eqnarray}
Here $\Delta \equiv \Delta m_{31}^2 L/(4 E)$ and $\hat{A} \equiv \pm
(2 \sqrt{2} G_F n_e E)/\Delta m_{31}^2$ with $G_F$ the Fermi coupling
constant and $n_e$ the electron density in matter. The sign of the
second term is positive for $\nu_{e} \rightarrow \nu_{\mu}$ or
$\nu_{\bar{\mu}} \rightarrow \nu_{\bar{e}}$ and negative for
$\nu_{\mu} \rightarrow \nu_{e}$ or $\nu_{\bar{e}} \rightarrow
\nu_{\bar{\mu}}$. The sign of $\hat{A}$ is determined by the sign of
$\Delta m_{31}^2$ and choosing neutrinos (factor $+1$) or
antineutrinos (factor $-1$).

This expansion clearly identifies $\deltacp$ as a genuine three-flavor 
effect in the second and third terms, which is double-suppressed by 
the mass hierarchy parameter $\alpha$ and $\sin 2 \theta_{13}$. Since 
$\sdm$ has turned out to be rather large within the LMA-allowed 
region~\cite{Bahcall:2004ut,Bandyopadhyay:2004da,Maltoni:2004ei}, 
it became possible to determine $\deltacp$, but the bottleneck of 
any CP measurement will certainly be the true value of $\stheta$. 
So far, $\stheta$ has been only restricted to $\stheta \lesssim 10^{-1}$ 
by the CHOOZ and Palo~Verde reactor data and recently also by the solar 
experiments~\cite{Maltoni:2003da}. If $\stheta>0$ cannot be established 
by any experiment, then CP effects in neutrino oscillations can not 
be detected. As we will discuss later, for $\stheta > 0$, the actual 
value of $\stheta$ will determine the strategy to measure $\deltacp$.

In principle, the second term in \equ{PROBMATTER} contains
the intrinsic information on $\deltacp$ close to the oscillation maximum,
whereas the third term proportional to
$\cos \deltacp \cdot \cos \Delta$ is suppressed close to the oscillation 
maximum (where $\sin \Delta \sim 1$). In particular, it is easy to see that
the probability difference $|P_{e \mu} - P_{\bar{e} \bar{\mu}}|$ (or 
$|P_{\mu e} - P_{\bar{\mu} \bar{e}}|$) in vacuum is just twice the second 
term in \equ{PROBMATTER}, which is often called ``CP-odd probability 
difference''. Therefore, it is well known that using both neutrinos and 
antineutrinos helps to extract the leptonic CP phase from \equ{PROBMATTER}.
However, matter effects enhance the neutrino channel and suppress the 
antineutrino channel (for a normal mass hierarchy). In addition, a direct 
measurement of $|P_{e \mu} - P_{\bar{e} \bar{\mu}}|$ is not possible, since 
in an actual experiment only the convolution of the probability, the flux, 
the cross sections, and detector efficiencies is measured. Each part of this
convolution may give rise to additional contributions to the 
`probability difference', especially cross sections and efficiencies might 
prove to make the use of a mere `probability difference' pointless. 
Furthermore, the appearance event rates are proportional to the full 
\equ{PROBMATTER}. Therefore, matter effects, matter density uncertainties, 
and the complicated parameter dependence in \equ{PROBMATTER} spoil the clean 
extraction of $\deltacp$. In particular, the complicated parameter dependence 
of the oscillation probability leads to correlations with 
$\stheta$~\cite{Cervera:2000kp,Burguet-Castell:2001ez} and multi-parameter
correlations~\cite{Huber:2002mx}, as well as to the $(\deltacp,
\theta_{13})$~\cite{Burguet-Castell:2001ez}, $\mathrm{sgn}(\Delta
m_{31}^2)$~\cite{Minakata:2001qm}, and
$(\theta_{23},\pi/2-\theta_{23})$~\cite{Fogli:1996pv} degeneracies,
\ie, an overall ``eight-fold'' degeneracy~\cite{Barger:2001yr}. In the
analysis, we will take into account all of these degeneracies. Note
however, that the $(\theta_{23},\pi/2-\theta_{23})$ degeneracy is not
present, since we always adopt for the true value of $\theta_{23}$ the
current atmospheric best-fit value $\theta_{23}=\pi/4$. Hence, we will
effectively deal with a ``four-fold'' degeneracy.

From the theoretical point of view, the second term in \equ{PROBMATTER}
is especially interesting for CP violation measurements, since the CP violating
effects are proportional to $\sin \deltacp$. However, only using information
close to the oscillation maximum, \ie, the $\sin \deltacp$-term,
leaves an intrinsic ambiguity between
$\deltacp$ and $\pi-\deltacp$. This ambiguity can be resolved with the
$\cos \deltacp$-term, as well as this term helps to disentangle
$\deltacp$ from $\stheta$. This means that for CP precision measurements
a contribution of this term is favorable. For example, neutrino
factories usually operate quite far off the oscillation maximum, which means
that there is some contribution of the $\cos \deltacp$-term anyway.
Note that for an experiment where the $\sin \deltacp$-term dominates, one
naturally expects a strong dependence of the CP precision on the true value of $\deltacp$
itself. In the high statistics dominated regime, the CP precision should
be best close to $0$ and $\pi$, since the derivative of $\sin \deltacp$ is
largest there, and worst to $\pi/2$ and $3/2 \, \pi$. This, of course,
can be changed by systematics and other effects.

Another interesting feature in \equ{PROBMATTER} is the condition
 $\sin( \hat{A} \Delta) = 0$ (not to mix up with $\hat{A} = 1$, which is the
 matter resonance condition). It makes all but the first term in 
\eq~(\ref{equ:PROBMATTER}) disappear and thus allows a clean measurement of 
$\sin^2 2 \theta_{13}$ and the sign of $\Delta m_{31}^2$ without correlations 
with the CP phase~\cite{Lipari:1999wy,Barger:2001yr}.
 This ``magic baseline''~\cite{Huber:2002uy,Huber:2003ak} condition is, for 
the first non-trivial solution, equivalent with $\sqrt{2} G_F n_e L = 2 \pi$ 
or, depending on the assumptions, $L \sim 7 \, 500 \, \mathrm{km}$. Its 
strength is the independence of all oscillation parameters, thus its value
is known {\it a priori}. In combination with a shorter baseline, it can be also
very efficient to access CP effects by resolving the intrinsic correlations
and degeneracies~\cite{Huber:2003ak}. We will show the obtainable precision
using this option, later on.

For the oscillation parameters, we use, if not stated otherwise, $\ldm
= 2.5 \cdot 10^{-3} \, \mathrm{eV}^2$, $\sin^2 2 \theta_{23} = 1$,
$\sdm = 7.0 \cdot 10^{-5} \, \mathrm{eV}^2$, and $\sin^2 2 \theta_{12}
= 0.8$. The numbers are similar to the ones from
\Refs~\cite{Fogli:2003th,Maltoni:2003da}.\footnote{However, the latest
  KamLAND results suggest a slightly higher value for
  $\sdm$~\cite{Bahcall:2004ut,Bandyopadhyay:2004da,Maltoni:2004ei},
  which were released after the calculations for this study have been
  performed (the overall calculation time was about three months).
  Since the CP effects are larger for larger values of $\sdm$, our
  results can be understood as the conservative limit.} In addition,
we assume a normal mass hierarchy, since it turns out that, though
there are quantitative differences, the assumption of an inverted mass
hierarchy does not produce excitingly new effects in CP
measurements~\cite{Huber:2002rs}. We only use values for
$\stheta$ below the current bound $\stheta \lesssim 0.1$~\cite{Maltoni:2003da}
and do not make any special assumptions
about $\deltacp$. However, we will show in some cases the results for
chosen selected values of $\deltacp$.

%%%%%%%%%%%%%%%%%%%%%%%%%%%%%%%%%%%%%%%%%%%%%%%%%%%%%%%%%%%%%%%%%%%%%%%%%%%%%
\section{Analysis methods}
\label{sec:analysis}
%%%%%%%%%%%%%%%%%%%%%%%%%%%%%%%%%%%%%%%%%%%%%%%%%%%%%%%%%%%%%%%%%%%%%%%%%%%%%

We now describe briefly the general analysis methods, the used experiments
and the computation of the CP coverage, which has to be done in an efficient
way.  In particular, the topology of the neutrino factory parameter space
including the $(\deltacp,\theta_{13})$-degeneracy makes the
computation of the CP coverage for very small values of $\stheta$,
\ie, a relatively flat topology, quite complicated.\footnote{Note that
  this degeneracy may not only appear for the best-fit solution, but
  also in the $\mathrm{sgn}(\ldm)$-degeneracy, leading (for maximal
  mixing) to a four-fold ambiguity.} Thus we will demonstrate how the
CP coverage, which is a highly condensed performance indicator, is
obtained from the (marginalized) $\Delta \chi^2$, how the degeneracies
are located, and what their effects are. Most of the shown examples
in this section will be computed for a neutrino factory, since this experiment implies
the highest level of sophistication.

%%%%%%%%%%%%%%%%%%%%%%%%%%%%%%%%%%%%%%%%%%%%%%%%%%%%%%%%%%%%%%%%%%%%%%%%%%%%
\subsection{The experiments and their simulation}
%%%%%%%%%%%%%%%%%%%%%%%%%%%%%%%%%%%%%%%%%%%%%%%%%%%%%%%%%%%%%%%%%%%%%%%%%%%%

In general, we use a three-flavor analysis of neutrino oscillations including
matter effects. The matter density profile is taken to be constant with 5\%
uncertainty, to take into account matter density uncertainties as well as
matter profile effects~\cite{Geller:2001ix,Ohlsson:2003ip,Pana}. The analysis
is performed with the $\Delta \chi^2$ method using the GLoBES
software~\cite{Huber:2004ka}. We take into account statistics, systematics,
correlations, and degeneracies, where the correlations originate
in the projection of the six-dimensional fit manifold onto the axis of the
parameter of interest~\cite{Huber:2002mx}. We use a local minimizer for this
projection, which means that one has to be especially careful to find all
degenerate solutions and not to miss any relevant local minimum. The
correlations account for the fact that an experiment (or combination of
experiments) cannot entirely resolve the intrinsic structure of
the oscillation probabilities, but effectively measures a combination of
the oscillation parameters. In principle, we assume that each experiment
(or combination of experiments) will provide the best measurement of the
leading atmospheric oscillation parameters at that time, which is coming
from the disappearance channels of the accelerator-based experiments. For the
solar parameters, we assume that the ongoing KamLAND
experiment will improve the errors down to a level of about $10\%$ on
each $\sdm$ and $\theta_{12}$~\cite{Gonzalez-Garcia:2001zy,Barger:2000hy}.

The best precision is obtained for neutrino factories, where we will in
most cases use the representative \NuFactII\ from \Ref~\cite{Huber:2002mx}.
In its standard configuration,
it uses $4 \, \mathrm{MW}$ target power ($5.3 \cdot
10^{20}$ useful muon decays per year), a baseline of $3 \, 000 \,
\mathrm{km}$, and a magnetized iron detector with a fiducial mass of
$50 \, \mathrm{kt}$. We choose a symmetric operation with $4 \,
\mathrm{yr}$ in each polarity. For comparison and reference, we use
the \JHFHK\ upgrade proposed in \Ref~\cite{Itow:2001ee} with a target
power of $4 \, \mathrm{MW}$, a baseline of $295 \, \mathrm{km}$, and a
water Cherenkov detector with a fiducial mass of $1 \, 000 \,
\mathrm{kt}$~\cite{Huber:2002mx}. It operates two years in the
neutrino running mode, and six years in the antineutrino running mode
to account for the lower antineutrino cross section.\footnote{For the
  analysis of $\deltacp$, it turns out that in most cases the optimal
  performance can be reached for almost equal numbers of total events
  in the neutrino and antineutrino operation modes. Thus, if we assume
  given total running time, the fraction of the optimal antineutrino
  running time is primarily determined by the cross section ratio
  between neutrinos and antineutrinos~\cite{Huber:2002rs}.}  In
addition, we compare in some cases with a scenario where we could be
in ten years from now, which corresponds to the scenario ``After ten
years'' from \Ref~\cite{Huber:2004ug}.\footnote{Here, we do not take
  into account the \MINOS , \ICARUS , and \OPERA\ experiments, since
  their contribution to the final result would be marginal.}  It
includes the two first-generation superbeams \JHFSK
~\cite{Itow:2001ee} and \NUMI ~\cite{Ayres:2002nm}, where the
simulation is described in \Ref~\cite{Huber:2002rs} and the parameters
used in \Ref~\cite{Huber:2004ug}. In summary, both correspond very
much to their standard scenarios as in the LOIs with a total running time
of five years in the neutrino running mode for each experiment.
However, for \NUMI , we use a baseline of $812 \, \mathrm{km}$ at an
off-axis angle of $0.72^\circ$, a target power of $0.43 \,
\mathrm{MW}$ ($4.0 \cdot 10^{20}$ protons on target per year) and a
low-Z-calorimeter as detector with a fiducial mass of $50 \,
\mathrm{kt}$~\cite{Ambats:2004js}.\footnote{Somewhat better results might be obtained with
  a TASD (Totally Active Scintillator Detector) with about half the
  detector mass.}  In addition, the scenario uses the large reactor
experiment \ReactorII\ introduced in \Ref~\cite{Huber:2003pm} to
disentangle $\theta_{13}$ and $\deltacp$,\footnote{Using a
  considerable amount of antineutrino running would also disentangle
  these two parameters.  However, longer running times would be needed
  to account for the lower antineutrino cross section, which means
  that this scenario is unlikely to fit into the coming ten years.}
which has a baseline of $1.7 \, \mathrm{km}$ and an integrated
luminosity of $8 \, 000 \, \mathrm{t \, GW \, yr}$. We call the
combined scenario \TEN .

The scenario \TEN\ is quite sophisticated, but it could be reached
within ten years from now and it illustrates how fast and how much
one could push for $\deltacp$ within moderate time scales.
Furthermore, we use \JHFHK\ as a very advanced superbeam upgrade
to discuss what one could achieve within the superbeams era. Finally,
and as the major part of this work, we investigate a standard
scenario \NuFactII\ for neutrino factories, which might be the
ultimate tool for $\deltacp$. A higher gamma $\beta$-beam
may have a comparable potential and take this role if technically
feasible, but this needs further study~\cite{betainprep}.

%%%%%%%%%%%%%%%%%%%%%%%%%%%%%%%%%%%%%%%%%%%%%%%%%%%%%%%%%%%%%%%%%%%%%%%%
\subsection{From $\boldsymbol{\Delta \chi^2}$ to the CP coverage}
%%%%%%%%%%%%%%%%%%%%%%%%%%%%%%%%%%%%%%%%%%%%%%%%%%%%%%%%%%%%%%%%%%%%%%%%

\begin{figure}[t!]
\begin{center}
  \includegraphics[width=16cm]{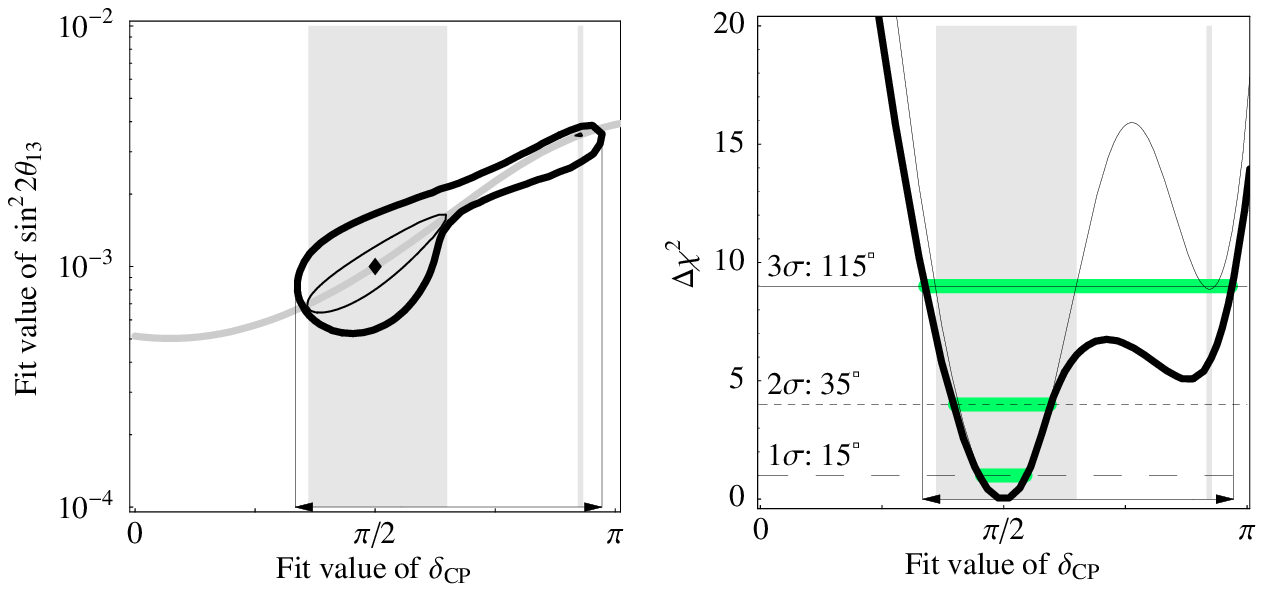}
\end{center}
\mycaption{\label{fig:cpcoverage} Correlation between $\deltacp$ and
  $\stheta$ for the simulated valued $\deltacp=\pi/2$ and
  $\stheta=0.001$ for \NuFactII .  The left panel shows the fit
  manifold in the $\deltacp$-$\stheta$-plane ($\Delta \chi^2 = 9$
  only), the right panel the projected $\Delta \chi^2$ onto the
  $\deltacp$-axis.  The thin curves refer to only taking into account
  the correlation between $\deltacp$ and $\stheta$, and the thick
  curves refer to taking into account the full multi-parameter
  correlation, \ie, also the correlations with the parameters which 
  are not shown.
  The shaded regions refer to the error in $\deltacp$ for the
  two-parameter correlation only, whereas the arrows mark the final
  error including the full correlation. The left panel shows in
  addition in gray the curve along which the minimum $\Delta \chi^2$
  in the $\stheta$-direction is found. In the right panel, the finally
  obtained CP coverage at the $1 \sigma$, $2 \sigma$, and $3 \sigma$ (1
  d.o.f.) levels is marked (thick horizontal lines) and given by
  numbers. Standard values are used for oscillation parameters which
  are not shown.}
\end{figure}

The CP coverage presents the information in a highly condensed manner, 
therefore it is useful to illustrate how it is obtained. In
\figu{cpcoverage}, we demonstrate this process in a simple example
without $\mathrm{sgn}(\ldm)$-degeneracy.  We compare the often used
picture in the $\deltacp$-$\stheta$-plane with the projection onto the
$\deltacp$-axis, where we use exactly the same scale on the horizontal
axis. In particular, we show the result in each panel with (thick
curves) and without (thin curves) correlations from parameters other
than $\stheta$.

First of all, the projection mechanism can be easily understood from
this figure. If one wants to know how precisely one can measure
$\deltacp$ in the left panel, fixing $\stheta$ will inevitably lead to
a too small error: The fact that we do at that time not know $\stheta$
more precisely than in this figure, leads to a larger error on
$\deltacp$. This two-parameter correlation is well-known to affect the
precision of $\deltacp$, and comes from the intrinsic structure of the
oscillation probabilities. One can include it in the analysis by
projection onto the $\deltacp$-axis, as it is done in the right panel.
We perform this projection by minimizing (marginalizing) $\Delta
\chi^2$ with respect to $\stheta$: If one takes the minimum $\Delta
\chi^2$ in the left panel in the direction of $\stheta$ for each fixed
value of $\deltacp$, one will find the minima along the gray curve.
The projected $\Delta \chi^2$ in the right panel is then nothing else
than the $\Delta \chi^2$ along the gray curve in the left panel.

Similarly to the two-parameter correlation, one can marginalize over
all of the not shown parameters (within the range allowed by external
data), leading to the full multi-parameter correlation. The
difference between only taking into account the two-parameter
correlation and the full correlation is illustrated by the difference
between the thin and thick curves in both panels of \figu{cpcoverage}.
One can easily see that, depending on the confidence level, the error
can be increased by more than 100\% by correlations other than
with $\stheta$ (unlike being small such as often indicated in the
previous literature). One can understand this mainly in terms of the
$(\deltacp,\theta_{13})$-degeneracy~\cite{Burguet-Castell:2001ez},
which, as a disconnected solution, may appear at a different place in
the $\deltacp$-$\theta_{13}$-plane (\cf, left panel, thin curve).
However, if one just fixes the other (not shown) oscillation
parameters, one does not account for the fact that the actual minimum may
lie slightly off the shown fit manifold section (\ie, plane) with
respect to these parameters. The full marginalization shows the full
beauty of this degeneracy.

The CP coverage in this example is obtained as the range of possible
fit values which fit the simulated value $\deltacp=\pi/2$. It is, for
three different confidence levels, marked and given in the right
panel. Note that there might be more than one fit region, which has
also to be included for the CP coverage by definition.  The right
panel clearly shows the non-Gaussian dependence of the CP coverage on
the confidence level. This dependence mainly originates in the role of
the $(\deltacp,\theta_{13})$-degeneracy, which may lead to connected
or disconnected degenerate solutions at the chosen confidence level.
Later, we will observe that this dependency can be amplified by the 
$\mathrm{sgn}(\ldm)$-degeneracy.

%%%%%%%%%%%%%%%%%%%%%%%%%%%%%%%%%%%%%%%%%%%%%%%%%%%%%%%%%%%%%%%%%%%%%%
\subsection{Localization of degeneracies}
%%%%%%%%%%%%%%%%%%%%%%%%%%%%%%%%%%%%%%%%%%%%%%%%%%%%%%%%%%%%%%%%%%%%%%

\begin{figure}[tp]
\begin{center}
  \includegraphics[height=17cm]{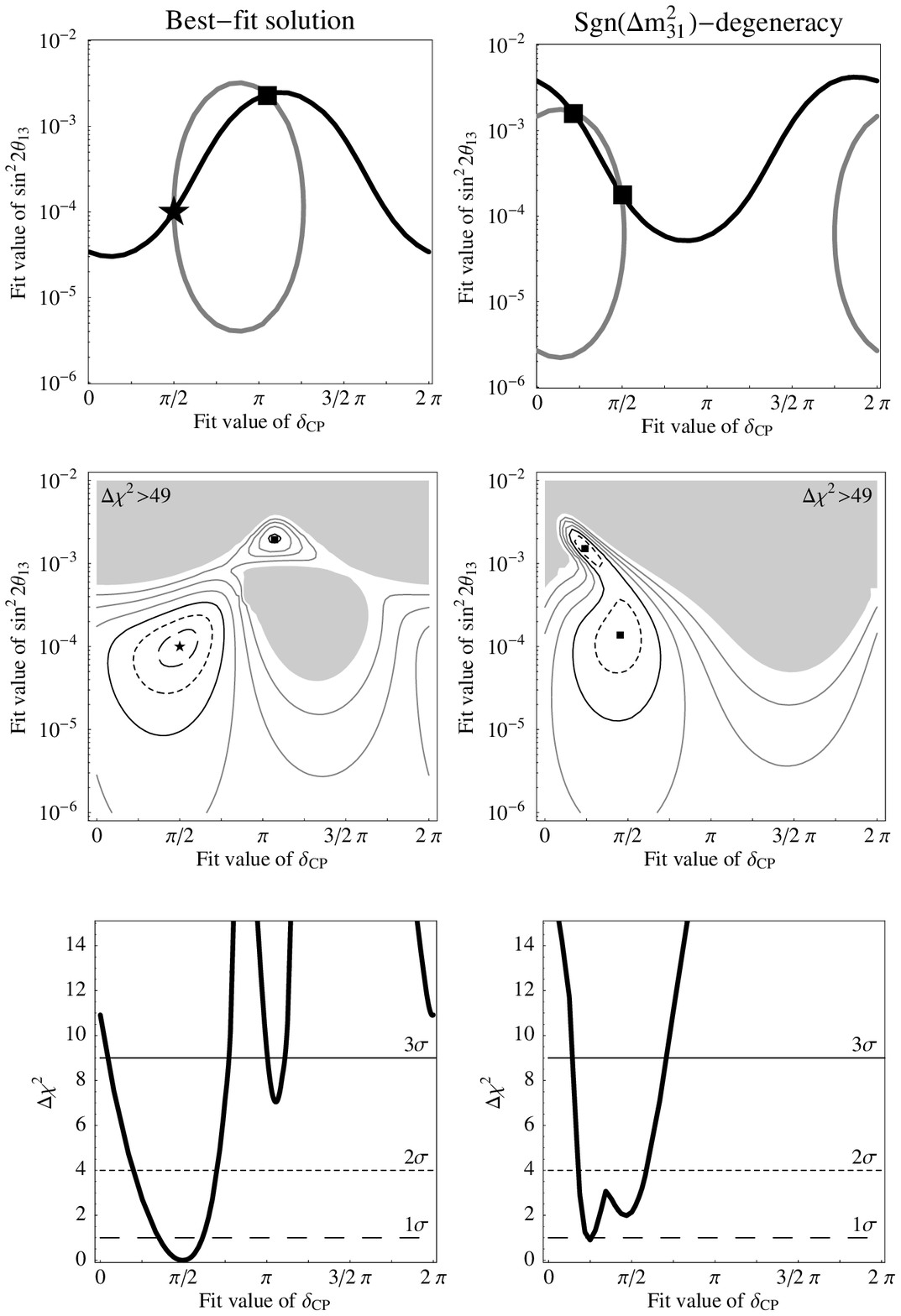}
\end{center}
\mycaption{\label{fig:degsummary} Degeneracy localization for the
  best-fit solution (left column) and $\mathrm{sgn}(\ldm)$-degeneracy
  (right column) for the simulated values $\deltacp = \pi/2$ and
  $\stheta = 0.0001$ and \NuFactII . The first row shows the total
  neutrino (black) and antineutrino (gray) constant rate curves as
  described in the text. The second row shows the projections of
  the fit manifolds onto the $\deltacp$-$\stheta$-plane, where the $1
  \sigma$ (long dashed), $2 \sigma$ (short dashed), $3 \sigma$ (solid
  black), $4 \sigma$, $5 \sigma$, and $6 \sigma$ (gray curves)
  contours (1 d.o.f.)  are shown. The gray-shaded region corresponds
  to $\Delta \chi^2>49$.  The third row shows the projection of the
  second row onto the $\deltacp$-axis, from which the final CP
  coverage is read off. In some of the panels, stars refer to the
  best-fit solution, and filled rectangles to the located
  degeneracies. Standard values are used for oscillation parameters
  which are not shown.
    }
\end{figure}

Besides the computation of the CP coverage for a specific solution in
parameter space, it is necessary to find all disconnected degenerate
solutions.  For maximal mixing, there are up to three disconnected
degenerate solutions besides the best-fit solution: one from the
$(\deltacp,\theta_{13})$-degeneracy, one from the
$\mathrm{sgn}(\ldm)$-degeneracy, and one from the mixed degeneracy.
Even if the $(\deltacp,\theta_{13})$-degeneracies were connected to
the original solutions at the chosen confidence level, we will see
later that it would in many cases be difficult to locate them
only with a local minimization method.

In order to locate all degenerate solutions, we use a method based on
the total event rates. To a first approximation, one
can use the total neutrino and antineutrino event rates for an
estimate of the positions of the degeneracies in the
$\deltacp$-$\stheta$-plane.  These degeneracies might later be
resolved by spectral information, or actually lie slightly off this
plane with respect to the fixed parameters, but it turns out that this
approach is rather efficient not to miss any degenerate solution.
Compared to the oscillation probabilities, the total event rates
already contain some energy-weighted information with respect to beam
flux, cross sections, and efficiencies. This means that the position
of the degeneracy can usually be more accurately determined than just using
oscillation probabilities (for which one particular value of $E$ had to
be chosen, which needed to be determined by a similar algorithm).

In particular, we first compute the total rates $N_0$ and $\bar{N}_0$
of the neutrino and antineutrino appearance channels of a given
experiment at the best-fit point (fixed simulated parameter values,
normal mass hierarchy).  Then we show the curves
$N(\deltacp,\stheta)_{|\ldm|>0}=N_0$ (black curve) and
$\bar{N}(\deltacp,\stheta)_{|\ldm|>0}=\bar{N}_0$ (gray curve) for
these constant neutrino and antineutrino rates as function of
$\deltacp$ and $\stheta$, as it is illustrated in the upper left panel
of \figu{degsummary}. By definition, both of these curves must go
through the best-fit point marked by the star, where they intersect.
It is now an interesting feature of the
$(\deltacp,\theta_{13})$-degeneracy that it is indistinguishable to
the best-fit solution on the total rate
level~\cite{Burguet-Castell:2001ez}. Thus, the second intersection
point in \figu{degsummary} (upper left panel) directly points to the
$(\deltacp,\theta_{13})$-degenerate solution.  Similarly, one can flip
the sign of $\ldm$ and show the curves for
$N(\deltacp,\stheta)_{|\ldm|<0}=N_0$ and
$\bar{N}(\deltacp,\stheta)_{|\ldm|<0}=\bar{N}_0$, where $N_0$ and
$\bar{N}_0$ are still the total rates at the best-fit point. The upper
right panel of \figu{degsummary} shows these curves for the inverted
mass hierarchy, leading to two more intersections, which means that we
have altogether located three (potential) degenerate solutions. Note
that the shape of the curves may change, and some curves may not even
intersect at all (except from the best-fit point). In the latter case,
the two closest points between the curves give a hint of the position
of the degeneracy.

As a next step, we start a local minimizer at each of the located
degeneracies (\ie, with $\deltacp$ and $\stheta$ of each degeneracy),
which minimizes the $\Delta \chi^2$ of the complete
experiment simulation (including spectral information) with respect to
all oscillation parameters. The resulting $(\Delta
\chi^2)_{\mathrm{min}}$ determines if a degenerate solution remains
below a chosen confidence level, or if it can be immediately resolved
by statistics, energy resolution \etc . The resulting position is the
actual position of the degeneracy taking into account the complete
statistical simulation. In the middle row of \figu{degsummary}, we
show the results from the complete simulation, where we project the
$\Delta \chi^2$ onto the $\deltacp$-$\stheta$-plane. In fact, one can
see that all located degeneracies are in this case present below the
$3 \sigma$ confidence level, and that their positions are very close
to the initial guesses from the total rate method.

Eventually, we show in the lower row of \figu{degsummary} the final
$\Delta \chi^2$ which is projected onto the $\deltacp$-axis. Note that for a
given set of simulated parameters, the final CP coverage is obtained
as union of all CP-ranges fitting the simulated value. Therefore, one
should think about an overlay between the left and right panels to
find the ranges which fit the simulated value of $\deltacp$. From the
comparison between the middle and lower rows of \figu{degsummary}, we
can demonstrate some more interesting properties of the topology for
small values of $\stheta$, too. First of all, as it can be inferred
from the left panels, in order to project onto the $\deltacp$-axis, it
is not sufficient to start the local minimizer at the best-fit value
of $\stheta$. In this specific case, one can easily image that
starting the minimizer at $(\pi, 10^{-4})$ could make it run back
towards smaller values of $\stheta$ instead of larger ones, where the
degeneracy is actually located. Therefore, the information from the
total rate approach is very useful to obtain an initial ``guess'' of
the value of $\stheta$ where the degeneracy is located at.  Another
interesting feature appears in the projected $\Delta \chi^2$ in the
lower right panel.  In this figure, the $\Delta \chi^2$ is actually
not a differentiable function, which seems to be very artificial at
the first sight. However, the comparison with the middle right panel
clearly illustrates the jump in $\stheta$ when moving from the left to
the right in $\deltacp$.  Again, only using some of the information
from the first row would not reveal the true structure of this
topological feature.

In summary, this total-rate-based approach turns out to be very
powerful for the localization of degeneracies and dealing with the
topology of neutrino factories for small values of $\stheta$.  In
principle, it can be also applied to the eight-fold degeneracy, where
the operation $\theta_{23} \rightarrow \pi/2 - \theta_{23}$ leads to
four more possible solutions.

%%%%%%%%%%%%%%%%%%%%%%%%%%%%%%%%%%%%%%%%%%%%%%%%%%%%%%%%%%%%%%%%%%%%%%%%%%
\subsection{Effects and interpretation of degeneracies}
%%%%%%%%%%%%%%%%%%%%%%%%%%%%%%%%%%%%%%%%%%%%%%%%%%%%%%%%%%%%%%%%%%%%%%%%%%

\begin{figure}[t!]
\begin{center}
  \includegraphics[width=12cm]{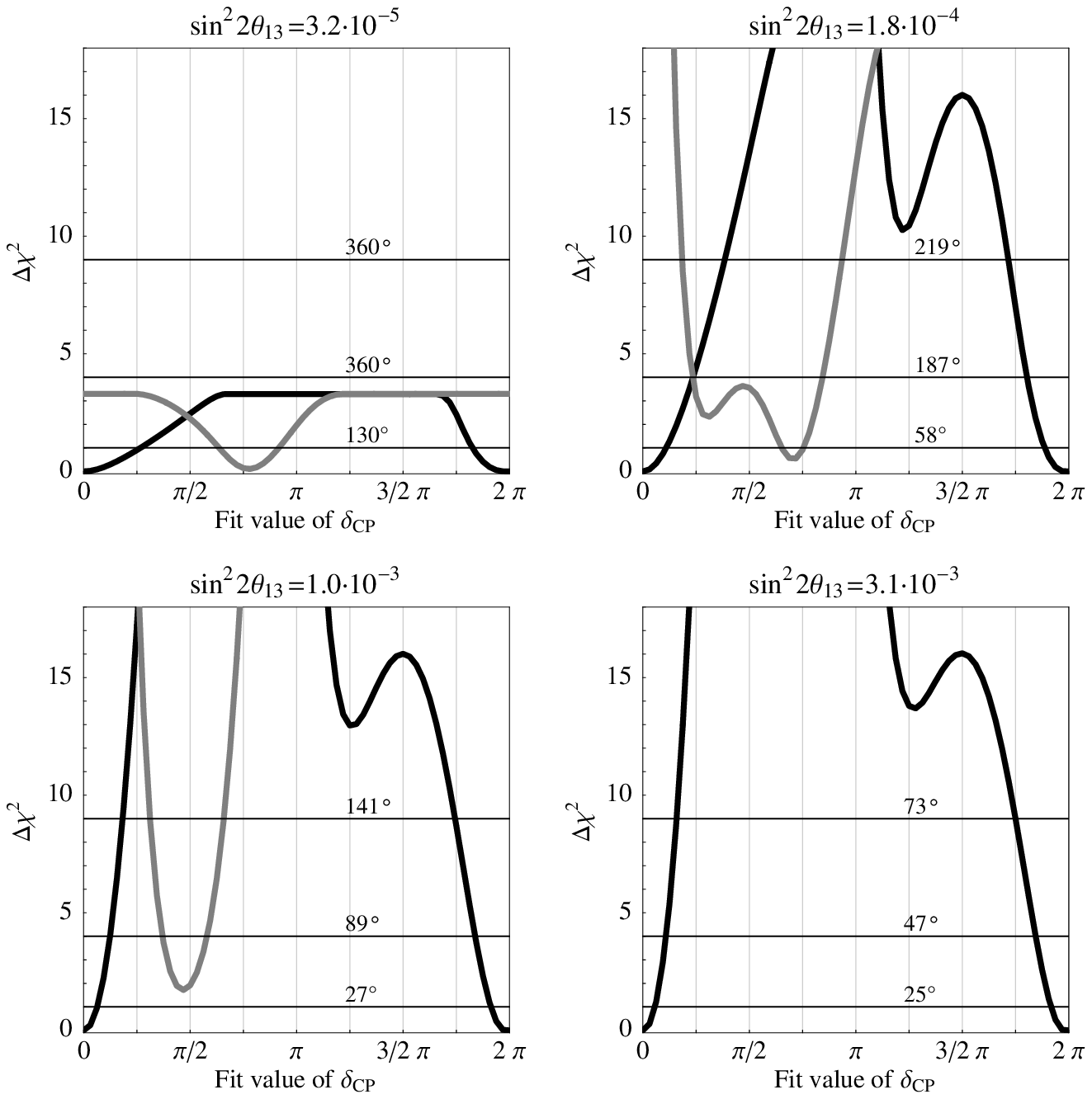}
\end{center}
\mycaption{\label{fig:degeffects} The projected $\Delta \chi^2$ for
  the simulated value $\deltacp=0$ and several (increasing) simulated values of
  $\stheta$ as given in the plot captions (\NuFactII ). In each
  figure, the original solution (black curve) and the
  $\mathrm{sgn}(\ldm)$-degeneracy (gray curve) is shown (degeneracy
  only shown if appearance below $3 \sigma$ confidence level). In
  addition, the final values of the CP coverage are given. For the
  other oscillation parameters, we use the standard values for this
  study.}
\end{figure}

In order to understand the effects of the degeneracies better, we show
an illustrative example in \figu{degeffects}. In this figure, the
projected $\Delta \chi^2$ is shown for the simulated value
$\deltacp=0$ and several increasing simulated values of $\stheta$ as
given in the plot captions. Both the original solutions (black curves)
and the $\mathrm{sgn}(\ldm)$-degeneracies (gray curves) are plotted.
For the smallest value of $\stheta$ (first panel), there is no
sensitivity to $\deltacp$ at the $2 \sigma$ and $3 \sigma$ confidence
levels. The $\Delta \chi^2$ is flat in a wide region, where
$\stheta=0$ acts as an attractor to the minimizer. At the $1 \sigma$
confidence level, however, the degeneracy appears at a different
position in $\deltacp$, which can easily double the CP coverage. Thus,
we learn that the degeneracy becomes especially important if it
introduces new values of $\deltacp$ compared to the original solution.
Note again that the final CP coverage is obtained as the union
of all regions fitting the true value, because any value within a
degenerate solution cannot be excluded at the chosen confidence level.

For somewhat larger values of $\stheta$ (second panel), the minimum of
the degeneracy is lifted.  In this example, there is also a
$(\deltacp,\stheta)$-ambiguity in both the original and
$\mathrm{sgn}(\ldm)$-degenerate solution, but only the one of the
$\mathrm{sgn}(\ldm)$-degeneracy affects the results below the $3
\sigma$ confidence level. In particular, it approximately doubles the
size of the $2\sigma$-allowed region. In this case, the original and
degenerate solutions hardly overlap, and the CP coverage of the
original solution is almost tripled at the $2 \sigma$ confidence
level. Because of the increasing overlap at larger $\Delta
\chi^2$-values, it does not change very much at the $3 \sigma$
confidence level, which is another indicator for a strongly
non-Gaussian behavior of the CP coverage.

For even larger values of $\stheta$ (right two panels), the
$\mathrm{sgn}(\ldm)$-degeneracy is lifted until it does not appear
below the $3 \sigma$ confidence level anymore.  The
$(\deltacp,\stheta)$-ambiguity in the $\mathrm{sgn}(\ldm)$-degeneracy
(mixed degeneracy) can be resolved by the better statistics in the
appearance channels. Reading \figu{degeffects} as a movie from left to
right (not to scale), it is obvious that problems with degeneracies
especially occur at an intermediate scale $10^{-4} \lesssim \stheta
\lesssim 5 \cdot 10^{-3}$, where the fit topology is rather flat.
Another important aspect in the effects of degeneracies is not so
obvious from this figure: For simulated values in the region of
$\deltacp \sim 3/2 \cdot \pi$ the degeneracy starts moving with
increasing $\stheta$. This has been illustrated in \fig~8 of
\Ref~\cite{Huber:2002mx}, where especially the position of the
degeneracy close to $\pi$ (``$\pi$ transit'') destroys the CP
violation sensitivity ($\stheta \sim 3 \cdot 10^{-4}$). Thus, the
structure of the topology strongly depends on the simulated parameter
values, which lead to a certain, in some cases complex configuration
and shape of original and degenerate solutions. Therefore, we cannot
expect to understand every feature of the CP coverage without analyzing
the respective topology.

%%%%%%%%%%%%%%%%%%%%%%%%%%%%%%%%%%%%%%%%%%%%%%%%%%%%%%%%%%%%%%%%%%%%%%%%%%
\section{From exclusion to high precision measurements of
  $\boldsymbol{\deltacp}$}
\label{sec:results1}
%%%%%%%%%%%%%%%%%%%%%%%%%%%%%%%%%%%%%%%%%%%%%%%%%%%%%%%%%%%%%%%%%%%%%%%%%%

\begin{figure}[t!]
\begin{center}
  \includegraphics[width=16cm]{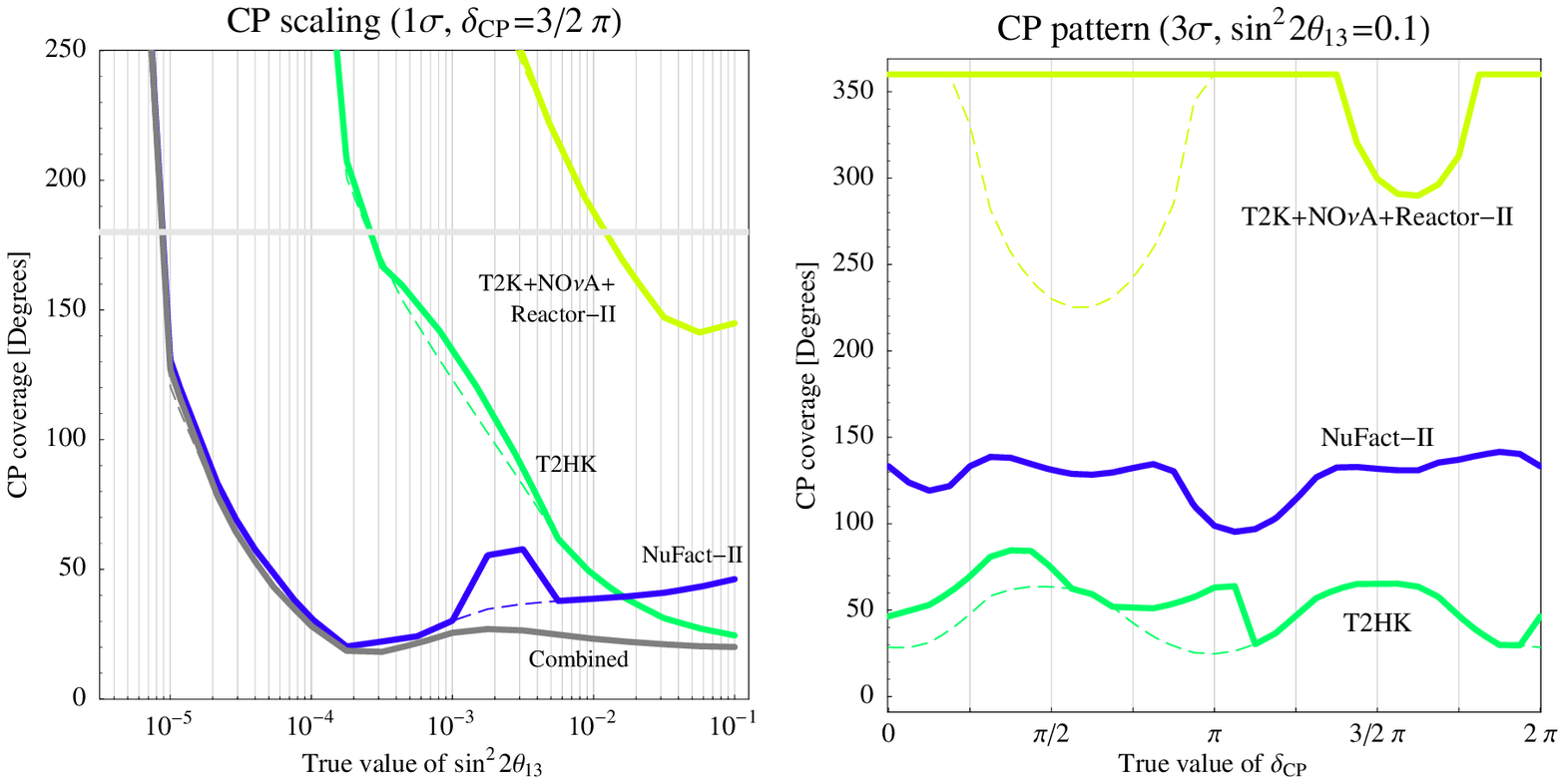}
\end{center}
\mycaption{\label{fig:allexps} The CP coverage as function of the true
  value of $\stheta$ (``CP scaling'') and as function of the true
  value of $\deltacp$ (``CP pattern'') for different experiments and
  parameters as given in the figures. The thin dashed curves refer to
  not taking into account the $\mathrm{sgn}(\ldm)$-degeneracy, \ie, if
  the mass hierarchy was known. The horizontal line (left panel) is an
  estimate for the range of values (below line), where CP violation
  measurements become possible. Standard values are used for the 
  oscillation parameters which are not shown. Note that the
  two figures use different confidence levels and different scales in
  the CP coverage.}
\end{figure}

In this section, we focus on the comparison of different classes of 
experiments, and we demonstrate the conceptual differences arising 
when going from one to the next generation of experiments. It is not
sufficient to show only the CP coverage for some discrete sets of 
parameter values in order to gain further insights. We therefore
define highly condensed performance indicators (\HCPI) as function 
of the most relevant parameter values. For the case of CP coverage 
it turns out that the dependence on the true (simulated) value of 
$\stheta$, and the true value of $\deltacp$ itself are the most 
interesting parameter dependencies. Though the latter seems to be 
strange at a first glance, it is exactly what was seen in
\Sec~\ref{sec:framework}, namely that the topology depends on the
value of $\deltacp$ actually realized by nature.  Thus, the correct 
way to interpret these functional dependencies is to read: ``If the 
true value of the parameter ... is ..., the CP coverage will be ... degrees.''.

In particular, we use the following two \HCPI:
\begin{description}
\item[CP scaling:] This is the CP coverage as function of the true
  value of $\stheta$ (for fixed simulated value of $\deltacp$).
  It is useful to compare the performance of different experiments
  as function of $\stheta$ and it is mainly constrained by the
  statistics of the appearance channels.
\item[CP pattern:] Defined as CP coverage as function of the true value
  of $\deltacp$ (for fixed simulated value of $\stheta$)~\cite{Winter:2003ye}. It is
  useful for a minimization of the risk with respect to the unknown
  true value of $\deltacp$ and it is mainly determined by the intrinsic
  structure of the oscillation probability in \equ{PROBMATTER}.
\end{description}
Examples for a CP scaling and a CP pattern can be found in \figu{allexps}.
With the GLoBES software~\cite{Huber:2004ka}, each of
these figures takes several days of computation time on a modern
computer, because about $10^9$ $\Delta \chi^2$'s have
to be evaluated.

An even more condensed performance indicator for CP
scalings can be found in \fig~19 of \Ref~\cite{Huber:2002mx}, which
takes the conservative case over all simulated values of $\deltacp$
(instead of fixing $\deltacp$).  One can easily imagine that, with
reasonably good resolution, such a figure has even an order of magnitude longer
computation time including all correlations and degeneracies than the ones
shown in this study.

%%%%%%%%%%%%%%%%%%%%%%%%%%%%%%%%%%%%%%%%%%%%%%%%%%%%%%%%%%%%%%%%%%%%%%%
\subsection{Comparison of experiment classes}
%%%%%%%%%%%%%%%%%%%%%%%%%%%%%%%%%%%%%%%%%%%%%%%%%%%%%%%%%%%%%%%%%%%%%%%

In \figu{allexps}, left panel, we compare different classes of
experiments, where the figure is shown at the $1 \sigma$ confidence
level for fixed true $\deltacp = 3 \pi/2$.  One class of these experiments
is the combined potential of the next generation of experiments under
optimistic assumptions. For the combination \TEN , where the large
reactor experiment could equally replaced by extensive antineutrino
running, already some exclusion of the parameter space of $\deltacp$
might be possible. In particular, for the shown parameter values, up
to about $210^\circ$ ($360^\circ-150^\circ$) of all possible values
could be disfavored at the $1 \sigma$ confidence level, where next
generation experiments are limited to the range $\stheta \gtrsim
10^{-3}$.  Note that a CP precision smaller than $180^\circ$ is a
necessary condition for maximal CP violation measurements, which means
that CP violation measurement with the shown combination will be
hardly possible at a useful confidence level (\cf, horizontal line in
left panel).  In addition, as one can see from \figu{allexps}, right
panel, a CP coverage smaller than $360^\circ$ is only available close
to $3 \pi/2$ (for which the left panel is shown).

As a representative for superbeam upgrades, we have shown \JHFHK\ in
both panels.  Since we assume a fiducial mass of $1 \, \mathrm{Mt}$,
it represents the upper possible limit for superbeams. As one can see
from \figu{allexps}, left panel, \JHFHK\ could give precise
information on $\deltacp$ in the range $\stheta \gtrsim 10^{-3}$: In
about this range, (maximal) CP violation measurements are possible
(\cf, \fig~18 of \Ref~\cite{Huber:2002mx}). In addition, for $\stheta \gtrsim 10^{-2}$
the information would be better than the one from the neutrino factory.
The right panel of \figu{allexps} clearly demonstrates that for large values of $\stheta$
very good CP precision measurements are possible regardless the true
value of $\deltacp$. In particular, it has been demonstrated in
\Refs~\cite{Huber:2002mx,Ohlsson:2003ip} that for large values of
$\stheta$ neutrino factories can be highly affected by matter density
uncertainties, which means that the superbeam upgrade could actually
be more competitive.

On the end of very small values of $\stheta$, neutrino factories could
give information on $\deltacp$ down to $\stheta \sim 10^{-5}$ (\cf,
\figu{allexps}, left panel). Especially in combination with \JHFHK ,
the CP coverage becomes very flat in $10^{-4} \lesssim \stheta
\lesssim 10^{-1}$. As far as the dependence on the true value of
$\deltacp$ is concerned (right panel), the dependence is rather flat for large values
of $\stheta$ because of the relatively broad energy spectrum. However,
we will see later, that this behavior changes for smaller values of
$\stheta$. Thus, in some sense, the parameter values for the two
panels in \figu{allexps} are chosen that both figures look very smooth
and show all investigated experiments. Nevertheless, \figu{allexps}
shows the quite generally applicable rule that the true value of
$\stheta$ determines the class of reasonable
experiments~\cite{Winter:2003st}.  This can be understood in terms of
the statistics in the appearance channel in \equ{PROBMATTER}: The
``signal'' in the second and third terms is simply proportional to
$\sin 2 \theta_{13}$, which means that the more events one has in this channel,
the smaller values of $\stheta$ allow the extraction of $\deltacp$.
Thus, there is no surprising result in the CP scalings.  The CP
patterns, however, will need further illumination in the next
subsections.

%%%%%%%%%%%%%%%%%%%%%%%%%%%%%%%%%%%%%%%%%%%%%%%%%%%%%%%%%%%%%%%%%%%%%%%%%%%%%%
\subsection{Characteristics of superbeams and the role of reactor experiments}
%%%%%%%%%%%%%%%%%%%%%%%%%%%%%%%%%%%%%%%%%%%%%%%%%%%%%%%%%%%%%%%%%%%%%%%%%%%%%%

Superbeams have some common characteristics, no matter if
next-generation experiments, or high-end superbeam upgrades. In
particular for the ones discussed here using the off-axis
technology~\cite{offaxis}, the beam spectrum becomes very narrow. The
$(\deltacp,\theta_{13})$-degeneracy does usually not appear in the
topology and causes no major problems. Neglecting the
$\mathrm{sgn}(\ldm)$-degeneracy, the dependence of the CP coverage on
the confidence level is therefore rather Gaussian. In the analysis,
there are hence no major complications expect from the
$\mathrm{sgn}(\ldm)$-degeneracy.

For these ``quasi-monochromatic'' beams, there is only little spectral
information, which means that many of their properties can be
understood on the oscillation probability or total rate level.  One
such interesting approach are bi-probability or bi-rate
graphs~\cite{Minakata:2001qm,Minakata:2003ca,Winter:2003ye}, which are
explicitly targeted towards understanding CP phase-dependent
properties. In particular, CP patterns for superbeams, such as in
\figu{allexps}, right panel, can be understood in terms of bi-rate
graphs~\cite{Winter:2003ye}, where different regimes can be
distinguished: For very high statistics, one obtains the typical
dependence as one would expect from the $\sin \deltacp$-term discussed
in \Sec~\ref{sec:framework} (for example, \figu{allexps}, right panel,
for \JHFHK , dashed curve).  For very poor statistics or large
systematical errors, the dependence on $\deltacp$ is reverted (\cf ,
\fig~4 of \Ref~\cite{Winter:2003ye}, left panel). For intermediate
statistics, this behavior can, in particular in connection with the
$\mathrm{sgn}(\ldm)$-degeneracy, lead to very complicated CP patterns
(\cf , \fig~4 of \Ref~\cite{Winter:2003ye}, middle panel). In
addition, the $\mathrm{sgn}(\ldm)$-degeneracy affects superbeam
measurements especially in the first and second quadrants, which means
that superbeam measurements are usually best between $\pi$ and $2
\pi$~\cite{Minakata:2001qm,Winter:2003ye}. To summarize, the
dependence on $\deltacp$ can be understood, though it might be
complicated. For \JHFHK , the strongest amplitude can be found for
$\stheta \lesssim 10^{-3}$, which can vary between $150^\circ$ and
$360^\circ$ by more than a factor of two -- between CP violation
measurements possible and no information on $\deltacp$ at
all~\cite{Winter:2003ye}.

Future reactor experiments could for $\stheta \gtrsim 10^{-2}$ help to
resolve the correlation between $\stheta$ and $\deltacp$ by a
precision measurement of $\stheta$ instead of extensive antineutrino
running at superbeams~\cite{Minakata:2002jv,Huber:2003pm,Anderson:2004pk}.
As it has been demonstrated in \Ref~\cite{Winter:2003ye} (\cf,
\fig~3), there are no major qualitative differences for the CP
patterns using antineutrino running or a large reactor experiment.
Therefore, the contribution of reactor experiments to CP measurements
can be understood in a similar way as the antineutrino running mode at
a superbeam.

%%%%%%%%%%%%%%%%%%%%%%%%%%%%%%%%%%%%%%%%%%%%%%%%%%%%%%%%%%%%%%%%%%%%%%
\subsection{Characteristics of neutrino factories}
%%%%%%%%%%%%%%%%%%%%%%%%%%%%%%%%%%%%%%%%%%%%%%%%%%%%%%%%%%%%%%%%%%%%%%

\begin{figure}[t!]
\begin{center}
  \includegraphics[width=15cm]{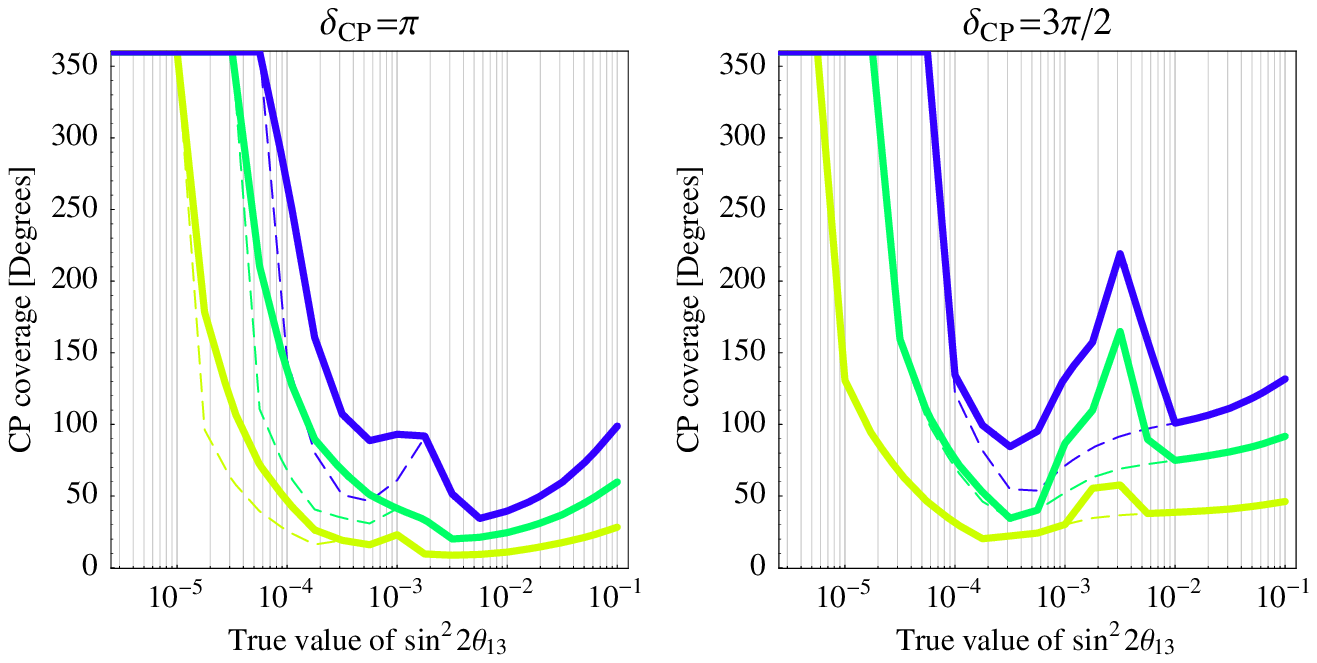}
\end{center}
\mycaption{\label{fig:nufactscalings} CP scalings for \NuFactII\ and
  two different selected values of $\stheta$ as given in the plot
  captions. The different curves in different colors correspond to the
  $1\sigma$, $2 \sigma$, and $3 \sigma$ confidence levels (from the
  lowest to the highest), where the dashed curves correspond to not
  taking into account the $\mathrm{sgn}(\ldm)$-degeneracy.  For the
  not shown oscillation parameters, we use the standard values for
  this study.}
\end{figure}

\begin{figure}[t!]
\begin{center}
  \includegraphics[width=15cm]{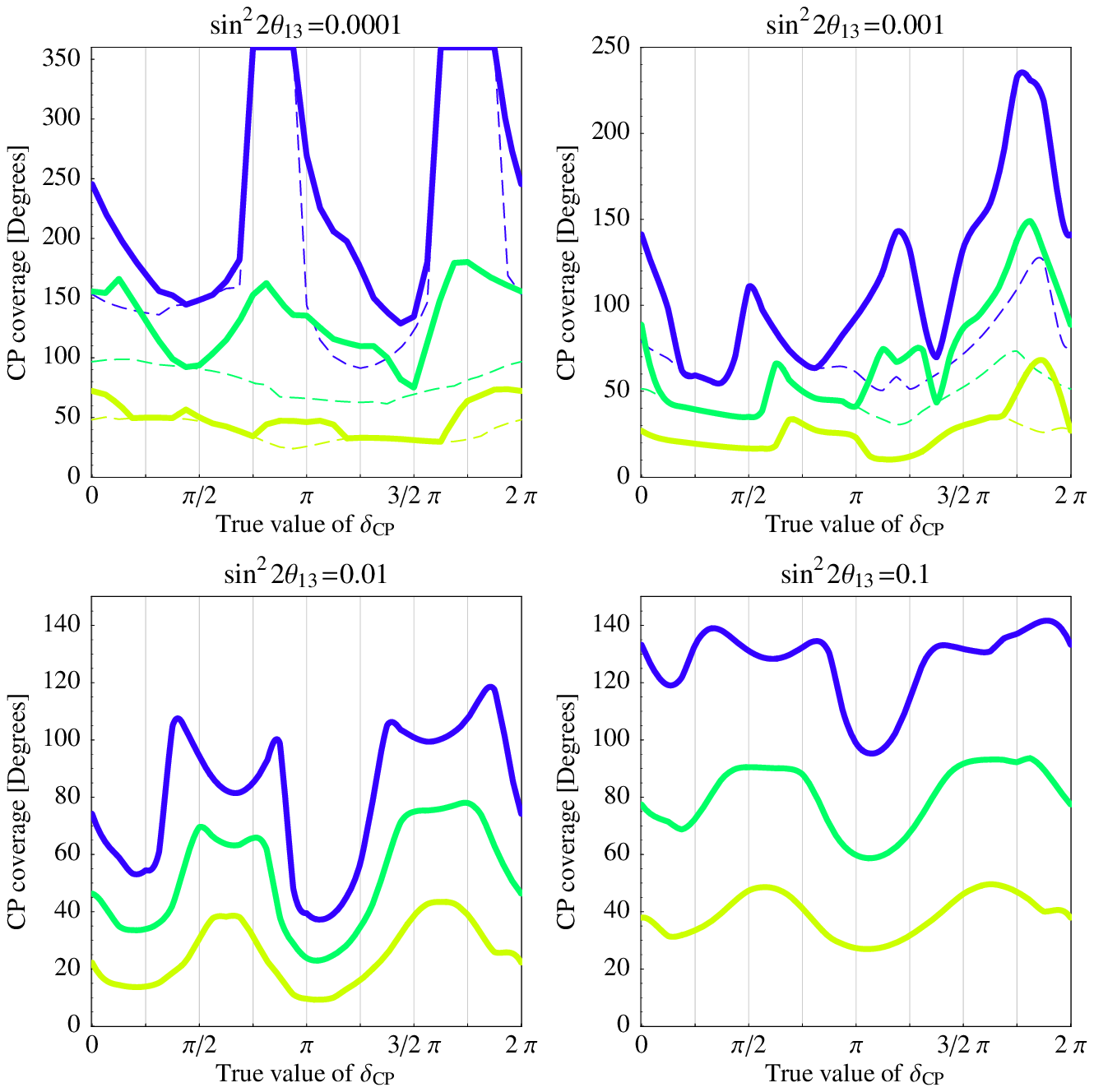}
\end{center}
\mycaption{\label{fig:nufactpatterns} CP patterns for \NuFactII\ and
  different values of $\stheta$ as given in the plot captions. The
  different curves in different colors correspond to the $1\sigma$, $2
  \sigma$, and $3 \sigma$ confidence levels (from the lowest to the
  highest), where the dashed curves correspond to not taking into
  account the $\mathrm{sgn}(\ldm)$-degeneracy.  For the not shown
  oscillation parameters, we use the standard values for this study.}
\end{figure}

As we have illustrated in \Sec~\ref{sec:analysis}, the analysis of
neutrino factories is much more complicated than the one for
superbeams because of the complicated topology. In particular, we can
not only expect a strong dependence on $\stheta$ or $\deltacp$ itself,
but also on the confidence level because of lifted (in CL) or moving
(in $\deltacp$) degeneracies and the importance of the relative
position to the best-fit solution.  In addition, neutrino factories
carry very good spectral information, which means that a simple
interpretation in terms of bi-rate graphs is not possible. However, we
will demonstrate how one can understand certain features and
limiting cases.

Let us first of all discuss the CP scaling for two selected, very
different cases in \figu{nufactscalings}. Obviously, the shape of
these scalings strongly depends on the confidence level and the true
value of $\deltacp$. In particular, we need to distinguish
irregularities in the dashed curves (no
$\mathrm{sgn}(\ldm)$-degeneracy), which are caused by the
$(\deltacp,\theta_{13})$-degeneracy, and irregularities which are only
present in the thick curves, which are caused by the
$\mathrm{sgn}(\ldm)$- or mixed degeneracies. If we only concentrate on
the $3\sigma$ curves (dark curves), we find that there are major
irregularities caused by the $\mathrm{sgn}(\ldm)$- or mixed
degeneracies in both cases. However, for $\deltacp=\pi$ (left panel),
also the thick curve is not smooth, which is due to the
$(\deltacp,\theta_{13})$-degeneracy being present around $\stheta
\lesssim 10^{-3}$.  For the case $\deltacp=3 \pi/2$ (right panel), the
mixed degeneracy (\ie, the $(\deltacp,\theta_{13})$-ambiguity in the
$\mathrm{sgn}(\ldm)$-degeneracy) doubles the CP coverage close to $2.5
\cdot 10^{-3}$. In general, degeneracies affect mainly the higher
confidence levels, since they usually disappear below a specific value
of $\Delta \chi^2$.  In addition, note that these interpretations
cannot be solely made from \figu{nufactscalings}, which means that one
has to look at less condensed information.

The CP patterns for \NuFactII\ are shown in \figu{nufactpatterns} for
different values of $\stheta$. As we have indicated above, their
interpretation is much more complicated than for superbeams because of
spectral information. First of all, note that there can be up to a
factor of five difference between the smallest and largest values of
the CP coverage, such as for $\stheta = 0.001$ ($3 \sigma$ CL). This
alone is a major hint that one should always compare the complete CP
patterns of different experiments and not only the results for
different selected values of $\deltacp$\footnote{In other words:
One can more or less always choose $\deltacp$ in such a way that some
experiment is better than the other one.}.
Compared to superbeams, we find similar principle cases:
\begin{description}
\item[High-statistics-dominated region] (large $\stheta$): The best
  performance can be achieved where the $\sin \deltacp$-term in
  \equ{PROBMATTER} has the steepest slope, \ie, close to $0$ and
  $\pi$. This can be clearly seen for $\stheta=0.1$, $1 \sigma$ CL.
  For larger confidence levels, however, the broad spectrum makes the
  dependence on $\deltacp$ flatter.
\item[Degeneracy-dominated region]
  At intermediate values of $\stheta \simeq 10^{-3}$ the four terms
  in \equ{PROBMATTER} have all approximately the same size,
  hence the problem of disentangling them is worst for those values
  of $\stheta$. Thus the relative position and appearance of the degeneracies
  is the major impact factor describing the CP pattern and can lead to
  complicated structures.
\item[Low-statistics-dominated region] (small $\stheta$): The behavior
  is inverted (best performance close to $\pi/2$ and $3 \pi/2$)
  because of the relatively large statistical errors, which can be
  understood in terms of bi-rate graphs (compare $3 \sigma$ curve for
  $\stheta = 0.0001$ in \figu{nufactpatterns} to the corresponding
  \fig~4 for \JHFHK\ of \Ref~\cite{Winter:2003ye}, left panel). The
  spectral information is in this case a subleading effect.
\end{description}

Especially, the second case of the degeneracy-dominated regime is very
interesting.  Analyzing the $3 \sigma$ curve of \figu{nufactpatterns}
for $\stheta=0.001$ (with the help of less condensed information), the
different peaks can be explained as follows: The first peak close to
$\deltacp=\pi/2$ is an effect of the
$(\deltacp,\theta_{13})$-degeneracy present for values of $\deltacp$
in this range. The peak close to $5 \pi/4$ comes from the
$\mathrm{sgn}(\ldm)$-degeneracy. This degeneracy, however, moves as
function of $\deltacp$ and overlaps the best-fit region for $11 \pi/8$
almost exactly, which leads to the sharp minimum.  For larger values
of $\deltacp$, not only the $\mathrm{sgn}(\ldm)$-degeneracy becomes
effective again (because it moves away from the original solution),
but also the $(\deltacp,\theta_{13})$-degeneracy in the original
solution and the same in the $\mathrm{sgn}(\ldm)$-degeneracy (mixed
degeneracy). Close to $7 \pi/4$ the overall four-fold degeneracy is
present with maximum non-overlap in $\deltacp$-space, leading to the
extremely high peak. One important consequence of the moving
degeneracy close to $\deltacp=3 \pi/2$ (\cf, \fig~8 of
\Ref~\cite{Huber:2002mx}) is that the neutrino factory is highly
affected by degeneracies in the third and fourth quadrants. This
behavior is very complementary to the one of the superbeams, which are
mainly affected in the first and second quadrants.  This explains the
synergy between superbeams and neutrino factories in certain regions
of the parameter space~\cite{Burguet-Castell:2002qx}.

%%%%%%%%%%%%%%%%%%%%%%%%%%%%%%%%%%%%%%%%%%%%%%%%%%%%%%%%%%%%%%%%%%%%%%%%%%
\section{Synergies in future high-precision measurements}
\label{sec:results2}
%%%%%%%%%%%%%%%%%%%%%%%%%%%%%%%%%%%%%%%%%%%%%%%%%%%%%%%%%%%%%%%%%%%%%%%%%%

\equ{PROBMATTER} shows that the actual value of $\stheta$ is the missing
key parameter for any measurement of $\deltacp$. We will therefore
organize the discussion of synergies for precision measurements of
$\deltacp$ by the true value of $\stheta$, \ie , the value which is
actually extracted from experiments (and corresponds to the simulated
value for simulations). Then we will discuss relevant synergies
for high precision measurements of $\deltacp$ using a more conceptual
line of reasoning.

%%%%%%%%%%%%%%%%%%%%%%%%%%%%%%%%%%%%%%%%%%%%%%%%%%%%%%%%%%%%%%%%%%%%%%%%%%
\subsection{The impact of the true $\boldsymbol{\stheta}$}
\label{sec:truesth13}
%%%%%%%%%%%%%%%%%%%%%%%%%%%%%%%%%%%%%%%%%%%%%%%%%%%%%%%%%%%%%%%%%%%%%%%%%

Unless a finite value of $\stheta$ is measured, each generation of
experiments will push the upper bound towards smaller values. One
might argue therefore, that a finite value of $\stheta$ has to be
established before it makes sense to discuss a measurement of $\deltacp$.
However, in fact these two parameters will in most cases (except from
reactor experiments) be simultaneously accessible, \ie, a positive
signal for $\stheta$ comes together with some information on
$\deltacp$.
Taking \figu{allexps}, left panel, as a rough estimate for the
sensitivity to $\deltacp$, we find the following approximate regions
in $\stheta$:
\begin{description}
\item[$\boldsymbol{\stheta \gtrsim 10^{-2}}$:] In this case,
  $\stheta>0$ will be very likely established by a superbeam or
  reactor experiment. The measurement of $\deltacp$ could then be
  pushed by existing superbeam experiments or upgraded versions (new
  baselines, new or larger detectors, more protons, longer running
  times, different $L/E$'s \etc).  The discussion of synergies then reduces to the level
  of superbeams and reactor
experiments~\cite{Wang:2001ys,Whisnant:2002fx,Barger:2002rr,Huber:2002rs,Minakata:2002jv,Minakata:2003ca,Migliozzi:2003pw,Huber:2003pm}.
  However, as one can also read off \figu{allexps} (left panel), for a
  high precision measurement of $\deltacp$, a very large superbeam
  upgrade, such as \JHFHK , is the choice to go for.  Already one such
  experiment, which encapsulates the synergy between the neutrino and
  antineutrino running, could deliver excellent information on
  $\deltacp$.  On the other hand, from the point of $\deltacp$, a neutrino
  factory is not needed to obtain high precision results (maximal mixing
  assumed). In particular, as it is illustrated
  in \fig~6 of \Ref~\cite{Ohlsson:2003ip}, matter density
  uncertainties highly affect the CP precision measurement for large
  values of $\stheta$. Effectively, they act as a normalization
  uncertainty of the first term in \equ{PROBMATTER}, which makes it
  hard to extract the second and third ones.
\item[$\boldsymbol{10^{-3} \lesssim \stheta \lesssim 10^{-2}}$:] This
  is one of the most interesting ranges for this synergy discussion,
  because it goes beyond the reach of next generation experiments, but
  it is well within the reach of superbeam upgrades and other
  alternative technologies (such as silver channel measurements or
  $\beta$-Beams)~\cite{Barger:2002xk,Burguet-Castell:2002qx,Donini:2002rm,Autiero:2003fu,Asratyan:2003dp,Burguet-Castell:2003vv,Donini:2004hu,Yasuda:2004gu,Donini:2004iv}.
\item[$\boldsymbol{10^{-4} \lesssim \stheta \lesssim 10^{-3}}$:] In
  this region, neutrino factories are the top candidates, which, however are
  highly affected by degeneracies.  This essentially means that the
  discussion of synergies reduces to the one of two neutrino factory
  baselines, such as in
  \Refs~\cite{Burguet-Castell:2001ez,Huber:2003ak}. In particular, it
  has been demonstrated in \Ref~\cite{Huber:2003ak} that the
  combination 3000 km + 7500 km (``magic baseline'', \cf\
  \Sec~\ref{sec:framework}) is the optimal choice for $\stheta$ (and
  other measurements).  We will therefore test this configuration for
  precision measurements of $\deltacp$.  So far, the only other
  competitive technology for $\stheta \lesssim 10^{-3}$ could be
  high-gamma $\beta$-Beams~\cite{Burguet-Castell:2003vv}, but their
  physics potential has to be evaluated further~\cite{betainprep}.
  Note that the combination, for instance, with a superbeam would not
  help much below $\stheta \lesssim 10^{-3}$, as it can be seen from
  \figu{allexps}, left panel.
\item[$\boldsymbol{10^{-5} \lesssim \stheta \lesssim 10^{-4}}$:] Only
  neutrino factories (and maybe high-gamma $\beta$-beams) could access
  this range. As we will see later, we are now for CP precision
  measurements in the (poor) statistics dominated regime, which means
  that correlations and degeneracies become a secondary impact factor.
  The discussion of synergies is the obsolete and reduces to an
  optimization of the statistics.  One neutrino factory baseline would
  probably be sufficient in this case.
\item[$\boldsymbol{\stheta \lesssim 10^{-5}}$:] This case needs further
study. However, if one really wants to built an experiment if $\stheta$ has not been
found at a neutrino factory, it is likely that a substantial luminosity
increase will be needed.
\end{description}

%%%%%%%%%%%%%%%%%%%%%%%%%%%%%%%%%%%%%%%%%%%%%%%%%%%%%%%%%%%
\subsection{One versus two neutrino factory baselines}
%%%%%%%%%%%%%%%%%%%%%%%%%%%%%%%%%%%%%%%%%%%%%%%%%%%%%%%%%%%

\begin{figure}[t!]
\begin{center}
  \includegraphics[width=16cm]{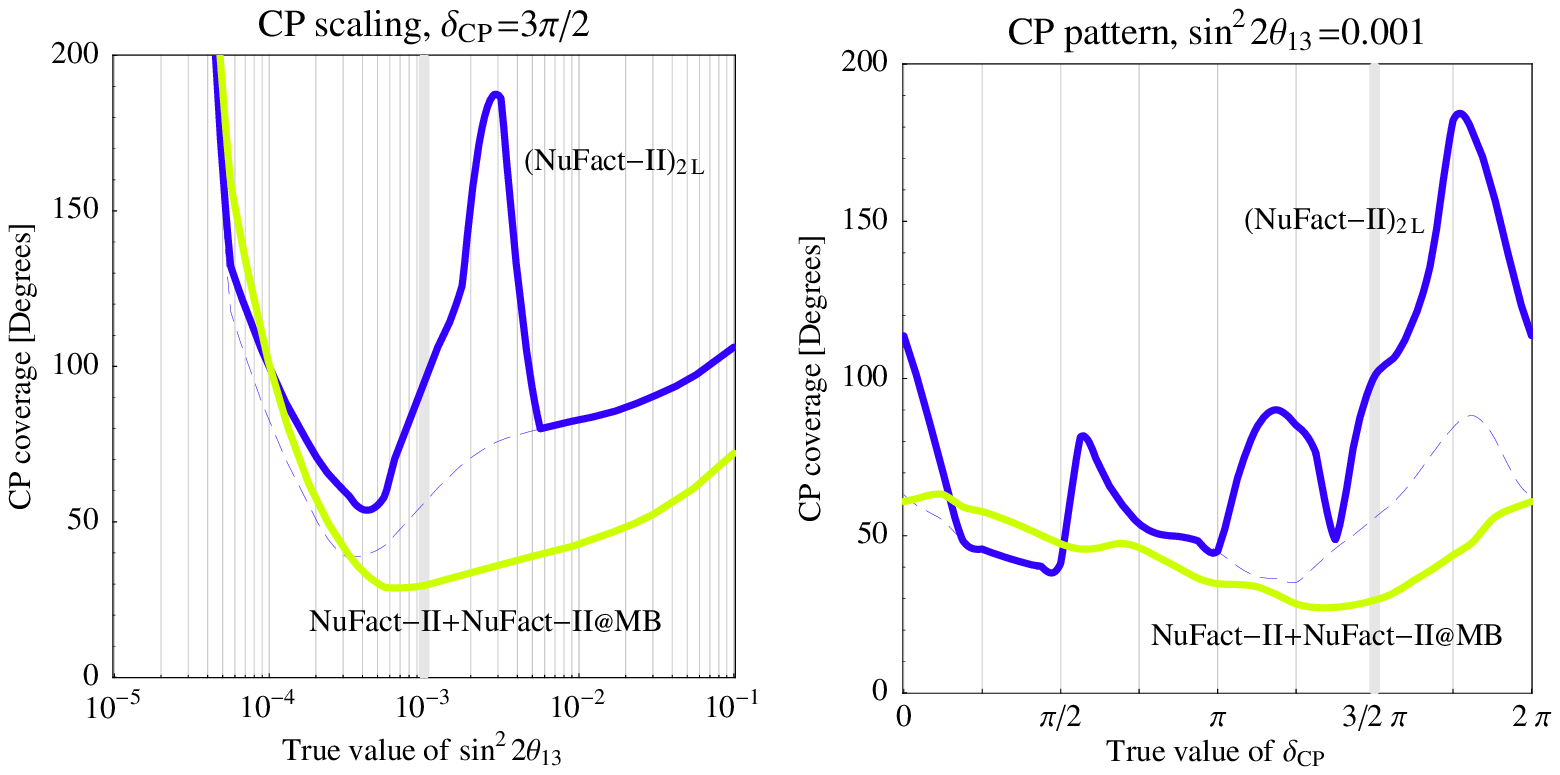}
\end{center}
\mycaption{\label{fig:magicsyn} The CP scaling and CP pattern for
  \NuFactII\ and the oscillation parameters given in the figure
  captions ($3 \sigma$ confidence level). The dark curves correspond
  to \NuFactII\ at $L=3 \, 000 \, \mathrm{km}$, but with double
  luminosity (2L). The light curves corresponds to \NuFactII\ with two
  detectors, which are located at $L=3 \, 000 \, \mathrm{km}$ and $L=7
  \, 500 \, \mathrm{km}$ (``magic baseline'' = MB) and operated with
  single luminosities each. This means that the product of useful muon
  decays times overall detector mass is equal in both shown cases.
  The dashed curves refer to not taking into account the
  $\mathrm{sgn}(\ldm)$-degeneracy.  The vertical lines mark the
  parameter values for which the plots correspond to each other.
  Standard values are used for the oscillation parameters  which are
  not shown.}
\end{figure}

Since a neutrino factory muon storage ring has (at least) two straight
sections, the option of a second detector should be natural when
planning the storage ring geometry. As it has been discussed in
\Sec~\ref{sec:framework}, the ``magic baseline'' of about $7 \, 500 \,
\mathrm{km}$ has turned out to be optimal for a ``clean'' measurement
of $\stheta$~\cite{Huber:2003ak} and very good for CP violation
measurements~\cite{Burguet-Castell:2001ez}. This means that one could
suspect very good synergy effects for high precision measurements of
$\deltacp$, since the correlation and degeneracy between $\deltacp$
and $\theta_{13}$ could be easily disentangled. As discussed in
\Sec~\ref{sec:truesth13}, the relevant parameter space region for this
option is $10^{-4} \lesssim \stheta \lesssim 10^{-2}$.

In \Ref~\cite{Huber:2002rs}, ``synergy'' between two or more
experiments/options has been defined as the ``extra gain [...] beyond
the simple addition of statistics''. Thus, in order to compare the
potential of one and two neutrino factory baselines, it is not enough
to simply add another baseline and compare it to the single baseline
option, \ie, one has to ``subtract'' the effect of the extra
statistics. We therefore compare in \figu{magicsyn} the potential of a
one- and two-baseline neutrino factory by using the double luminosity
for the single baseline option, which means that both options are
using the same product of useful muon decays times overall detector
mass. Thus, given a total amount of detector mass, it refers to the
question if one should put it all to $3 \, 000 \, \mathrm{km}$, or a
part of it (in our case half) to $7 \, 500 \, \mathrm{km}$.  As
performance indicators, we choose the CP scaling for $\deltacp = 3
\pi/2$ and the CP pattern for $\stheta$, since we know from the
discussion in \Sec~\ref{sec:results1} that these choices of parameters
represent the most critical regions within the parameter space. In
particular, it can be read off the right panel of \figu{magicsyn} that
$\deltacp=3 \pi/2$ is well within a region where degeneracies are
present, and it can be read off the left panel of \figu{magicsyn} that
$\stheta=0.001$ is in the critical range. Besides that,
$\stheta=0.001$ is in the middle of the parameter range of interest
here.

The CP scaling in \figu{magicsyn}, left panel, is an important
indicator to discuss the synergy between two neutrino factory
baselines in the complete parameter range of interest in $\stheta$. It
turns out that in the complete range $10^{-4} \lesssim \stheta
\lesssim 10^{-2}$ there is a real synergy between the $3\, 000 \,
\mathrm{km}$ baseline and the magic baseline. In particular, the
effect can be up to half of the parameter space for $\deltacp$. One
can also see from this panel that below $\stheta \lesssim 10^{-4}$ the
$3 \, 000 \, \mathrm{km}$ baseline alone is actually doing better,
which is the low-statistics-dominated region discussed in
\Sec~\ref{sec:truesth13}.

The CP pattern in \figu{magicsyn}, right panel, is the risk assessment
indicator with respect to the unknown true value of $\deltacp$. As one
can see from the curve for (\NuFactII )$_{\mathrm{2L}}$, there is a
range for the CP coverage from $40^\circ$ to $190^\circ$, \ie, almost
a factor of five. Not only can the minimum of this range reduced with
the magic baseline option, \ie, the performance at the best point, but
also the amplitude to about $40^\circ$ compared to $150^\circ$ before.
Thus, the risk of ending up with a very poor measurement of $\deltacp$
-- just because nature was not kind -- is substantially lowered.
However, it is important to note that there are regions in parameter
space where no synergy is needed because problems with correlations
and degeneracies are small. In particular, the single baseline option
does very well in the first quadrant, which means that the second
baseline would be a waste of resources in this case. Thus, we conclude
that one should definitively design the muon storage ring for these two baselines.
However, one may use a
staged approach with adding the second detector later.\footnote{Note
 that the muon storage ring tunnel for the $7 \,
  500 \, \mathrm{km}$ option might actually be a major cost factor
  compared to the second detector because of the strong decay tunnel
  slope. As well as the installation of the beam pipe and magnets at such
  a steep angle. Therefore, the staged approach might be a good solution.}

%%%%%%%%%%%%%%%%%%%%%%%%%%%%%%%%%%%%%%%%%%%%%%%%%%%%%%%%%%%%%
\subsection{Superbeam upgrade plus neutrino factory?}
%%%%%%%%%%%%%%%%%%%%%%%%%%%%%%%%%%%%%%%%%%%%%%%%%%%%%%%%%%%%%

\begin{figure}[t!]
\begin{center}
  \includegraphics[width=8cm]{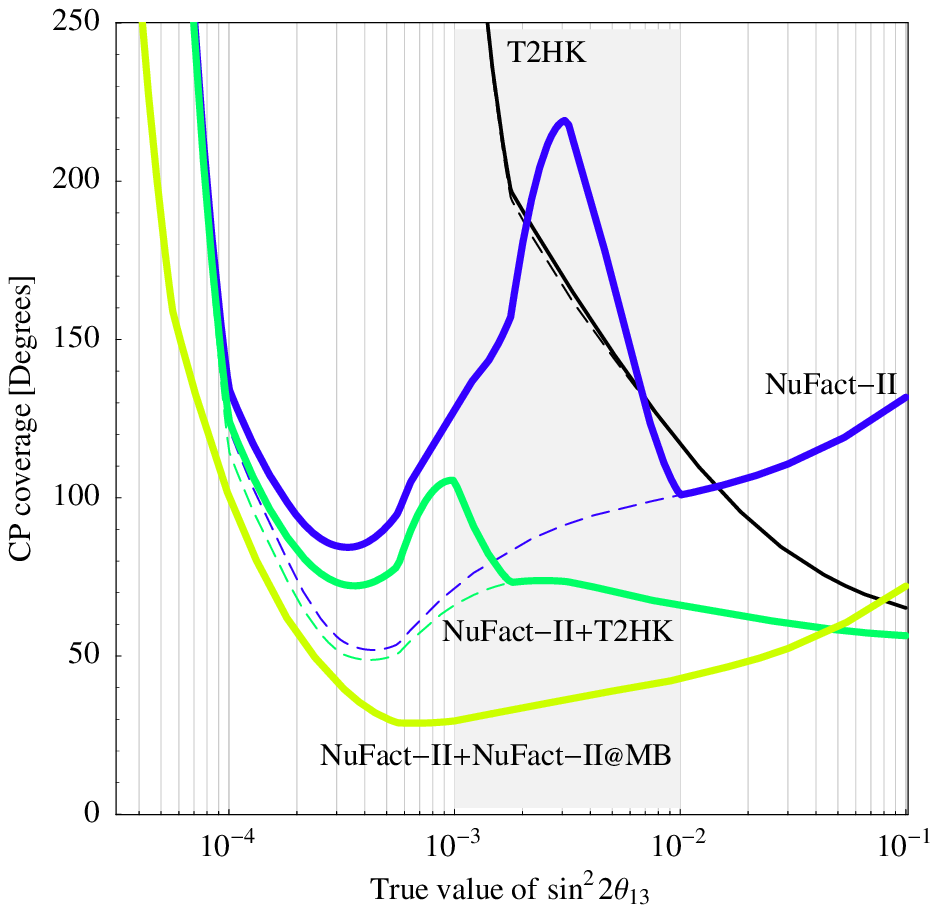}
\end{center}
\mycaption{\label{fig:superbeamsyn} The CP scaling for the experiments
  as given in the figure ($3 \sigma$ confidence level, $\deltacp = 3
  \pi/2$). The dashed curves refer to not taking into account the
  $\mathrm{sgn}(\ldm)$-degeneracy. The shaded region corresponds to
  the $\stheta$-range where synergy effects with the superbeam are
  present.  For the other oscillation parameters, we choose the
  standard values from this study. Details about the setups indicated
  by the labels can be found in the text.}
\end{figure}

As it has been pointed out in \Sec~\ref{sec:results1}, superbeams and
neutrino factories have a complementary behavior in $\deltacp$-space
if $\deltacp$ is in the third or fourth quadrants. Therefore, it is a
natural question to ask for the synergy between superbeam upgrades and
neutrino factories~\cite{Burguet-Castell:2002qx}. In this case, it is
not obvious how to show synergy effects, \ie, how to distinguish
between addition of statistics and real gain in physics potential. In
order to test this synergy, we show in \figu{superbeamsyn} the CP
scalings for \NuFactII , \JHFHK , \NuFactII +\JHFHK , and \NuFactII +
\NuFactII @MB ($L=3 \, 000 \, \mathrm{km}$ plus $L=7 \, 500 \, \mathrm{km}$
at ``Magic Baseline'') for $\deltacp = 3 \pi/2$.
First of all, one can see from
this figure that the combination of \NuFactII\ and \JHFHK\ indeed
performs much better than the individual experiments. In fact, in
comparison with (\NuFactII)$_{\mathrm{2L}}$, one can show that there
is a real synergy effect in $10^{-3} \lesssim \stheta \lesssim
10^{-2}$. However, there is no real gain below $\stheta \lesssim
10^{-3}$ except from adding a bit of statistics, which confirms that
superbeam upgrades would not help much below $\stheta \lesssim
10^{-3}$.  As far as the CP pattern is concerned, one can show that
the synergy effect is restricted to the third and fourth quadrants,
which means that the combination with a superbeam upgrade only helps
in a part of the parameter space.

One can now ask the question if one should have a superbeam upgrade or
a second neutrino factory baseline instead. If the assume that the
neutrino factory is necessary for the gray-shaded range in
\figu{superbeamsyn}, \ie, that it will be built anyway, it could be
fair to compare the additional effort for the magic baseline
(including $50 \, \mathrm{kt}$ detector) with the megaton water
Cherenkov detector necessary for \JHFHK . Comparing the respective
curves in \figu{superbeamsyn}, the magic baseline option is about a
factor of two better in the discussed $\stheta$-range. Since there is
no real synergy effect between neutrino factory and superbeam upgrade
below $\stheta \lesssim 10^{-3}$ anymore, this ratio increased to
about a factor of three there. Thus, if $\stheta$ turns out to be
smaller than $10^{-3}$, the superbeam upgrade would not help.

%%%%%%%%%%%%%%%%%%%%%%%%%%%%%%%%%%%%%%%%%%%%%%%%%%%%%%%%%%%%%%%%%%%%%%%%%
\subsection{Other cases: Alternative technologies or non-maximal mixing}
%%%%%%%%%%%%%%%%%%%%%%%%%%%%%%%%%%%%%%%%%%%%%%%%%%%%%%%%%%%%%%%%%%%%%%%%%

The last three subsections have focused on reasonably well-established
technologies and the current best-fit case of maximal atmospheric
mixing.  In this subsection, we will qualitatively discuss possible
alternatives to these scenarios.

Except from different detector technologies, the most interesting
alternative beam could be a $\beta$-beam~\cite{Zucchelli:2002sa,Bouchez:2003fy}.
In particular, if a higher gamma $\beta$-beam~\cite{Burguet-Castell:2003vv}
is feasible, then it could be a real competitor to neutrino factories.
Another challenging conceptual alternative for synergy effects is the
$\nu_\tau$ appearance ``silver'' channel~\cite{Donini:2002rm,Autiero:2003fu}.
We do not analyze these options quantitatively in this study, since
it is premature to anticipate the most promising direction on the
relevant time scales. In addition, there are different technological
questions to be clarified, and the analysis would therefore go
far beyond the scope of this work.
However, for the precision of $\deltacp$, we can qualitatively
formulate the requirements of a competitor to a neutrino factory from
\figu{magicsyn} (including maximal potential, \ie, ``magic baseline''
option):
\begin{quote}
The {\em alternative} configuration should be able to measure
  $\deltacp$ in the {\em full range} $10^{-4} \lesssim \stheta
  \lesssim 10^{-2}$ and $0 \lesssim \deltacp < 2 \pi$ to about
  $50^\circ$ at the $3 \sigma$ confidence level (including all
  correlations and degeneracies).
\end{quote}
Similarly, if we assume that we have a neutrino factory with one
baseline, the requirement for synergy effects reads
\begin{quote}
The {\em complementary} experiment should be able to provide
  good information on $\deltacp$ in the {\em full range} $10^{-4}
  \lesssim \stheta \lesssim 10^{-2}$ and the third and fourth
  quadrants in $\deltacp$.
\end{quote}
Note that it is dangerous to compare two configurations for only
specific selected true values of $\stheta$ and $\deltacp$, since once
can always find regions in parameter space which make this
configuration appear to be very competitive. In addition, note that
it may not be necessary to have a complementary experiment at all
if an eariler measurement confirmed $\deltacp$ be in the first or
second quadrants.

Besides the technological options or alternatives, there are also
physics alternatives. Within the framework of three-flavor neutrino
oscillations, the most likely alternative might be a substantial
deviation from maximal mixing which would lead to the full
eight-fold degeneracy~\cite{Barger:2001yr}.\footnote{Note, however,
  that there has to be a substantial deviation from maximal mixing
  to separate the $\theta_{23}>\pi/4$ and $\theta_{23}<\pi/4$ fit
  regions and to change the discussion qualitatively.} In this case,
further options could be necessary to resolve this degeneracy.
Depending on the parameter region, possible options include large
reactor experiments or superbeams for $\stheta \gtrsim
10^{-2}$~\cite{McConnel:2004bd,Burguet-Castell:2002qx}, very long
baseline superbeam upgrades~\cite{Diwan:2004bt}, an additional
$\beta$-beam facility~\cite{Donini:2004hu}, or another neutrino
factory baseline. In each of these cases, the determination of the
competitive parameter space region needs further investigation.

Finally there are further reasons which justify different technologies
or channels. The $\nu_\tau$ appearance channel (silver channel)
might, for example, be the best candidate for unitarity tests,
since neutral current measurements at neutrino factories do not
seem to be promising high precision measurements~\cite{Barger:2004db}.
In the same spirit, interferences with lepton flavor violating
operators \cite{Johnson:1999ci,Gago:2001xg,Gonzalez-Garcia:2001mp,Huber:2001de,%
Ota:2001pw,Huber:2002bi,Ota:2004ti}, neutrino decay (see
\eg~\Ref~\cite{Lindner:2001fx}), or other effects might lead to a different strategy.

%%%%%%%%%%%%%%%%%%%%%%%%%%%%%%%%%%%%%%%%%%%%%%%%%%%%%%%%%%%%%%%%%%%%%%%%
\section{Summary and conclusions}
\label{sec:summary}
%%%%%%%%%%%%%%%%%%%%%%%%%%%%%%%%%%%%%%%%%%%%%%%%%%%%%%%%%%%%%%%%%%%%%%%%

% Motivation of CP precision/CP coverage: Why better than CP violation

In this study, we have discussed $\deltacp$ precision measurements,
where we have used the concept of CP coverage as performance indicator.
CP coverage represents the fraction of the remaining parameter space for
$\deltacp$, \ie, small values correspond to very high precisions,
large values to very poor measurements, and $360^\circ$ to no information
on $\deltacp$ at all. In comparison to CP violation measurements, the CP coverage
exploits all available information on $\deltacp$. A measurement of CP
violation will, for example, hardly be possible for next-generation
experiments, whereas some values of $\deltacp$ could already be excluded.
In addition, for high precision measurements, the detection of CP
violation might not be possible if the true value of $\deltacp$ is close
to a CP conserving value. The concept of the CP coverage carries therefore
more information than a discussion of CP violation alone.

\begin{table}[h!]
\begin{center}
\begin{tabular}{lrrrrrrrr}
\hline
Experiment/Combination &  \multicolumn{2}{c}{$\delta_{\mathrm{CP}}=0$} & \multicolumn{2}{c}{$\delta_{\mathrm{CP}}=\pi/2$} & \multicolumn{2}{c}{$\delta_{\mathrm{CP}}=\pi$} & \multicolumn{2}{c}{$\delta_{\mathrm{CP}}=3/2 \, \pi$}  \\
\hline
\\[-0.3cm]
\fbox{$\sin^2 2 \theta_{13}=0.1$} \vspace*{0.2cm} \\
\JHFSK +\NUMI +\ReactorII & $90^\circ$ & ($360^\circ$) & $108^\circ$ & ($360^\circ$) & $96^\circ$ & ($360^\circ$)  & $145^\circ$ & ($300^\circ$)  \\
\JHFHK & $10^\circ$ & ($46^\circ$) & $24^\circ$ & ($75^\circ$) & $7^\circ$ & ($63^\circ$) & $25^\circ$ & ($65^\circ$)  \\
\NuFactII & $38^\circ$ & ($133^\circ$) & $47^\circ$ & ($131^\circ$) & $28^\circ$ & ($99^\circ$) & $46^\circ$ & ($132^\circ$)  \\
\\
\fbox{$\sin^2 2 \theta_{13}=0.01$} \vspace*{0.2cm} \\
\JHFSK +\NUMI +\ReactorII & $259^\circ$ & ($360^\circ$) & $232^\circ$ & ($360^\circ$) & $360^\circ$ &  ($360^\circ$)  & $187^\circ$ & ($360^\circ$)  \\
\JHFHK & $41^\circ$ & ($134^\circ$) & $58^\circ$ & ($132^\circ$) & $42^\circ$ & ($162^\circ$) & $47^\circ$ & ($117^\circ$)  \\
\NuFactII & $24^\circ$ & ($74^\circ$) & $31^\circ$ & ($94^\circ$) & $11^\circ$ & ($40^\circ$) & $39^\circ$ & ($101^\circ$)  \\
\\
\fbox{$\sin^2 2 \theta_{13}=0.001$} \vspace*{0.2cm} \\
\JHFSK +\NUMI +\ReactorII & $360^\circ$ & ($360^\circ$) & $360^\circ$ & ($360^\circ$) & $360^\circ$ & ($360^\circ$)  & $296^\circ$ & ($360^\circ$)  \\
\JHFHK & $139^\circ$ & ($360^\circ$) & $124^\circ$ & ($360^\circ$) & $360^\circ$ & ($360^\circ$) & $135^\circ$ & ($360^\circ$)  \\
\NuFactII & $27^\circ$ & ($141^\circ$) & $17^\circ$ & ($111^\circ$) & $23^\circ$ & ($93^\circ$) & $30^\circ$ & ($133^\circ$)  \\
\hline
\end{tabular}
\end{center}
\mycaption{\label{tab:cptable} The CP coverage for different simulated values
  of $\deltacp$ (columns) and $\stheta$ (row groups) for the indicated
  experiments or combinations of experiments.
CP coverage represents the fraction of the remaining parameter space for
$\deltacp$, \ie, small values correspond to very high precisions,
large values to very poor measurements, and $360^\circ$ to no information
on $\deltacp$ at all. The numbers are given at the
  $1 \sigma$ confidence level ($3 \sigma$ confidence level). For the other
  oscillation parameters, we use the standard values in this study. The
  definition of the assumed experimental scenarios is given in the text.}
\end{table}

So far there exist many, very different quoted values for the precision
of $\deltacp$ in the literature which are based on special values of
an assumed CP phase $\deltacp$. As one can see from \Tab~\ref{tab:cptable},
these different values are justified for these special values of
$\deltacp$. Depending on parameter values and confidence level
one can, for example, easily obtain any value between $7^\circ$
(very precise measurement) and $360^\circ$ (no
information on $\deltacp$) for \JHFHK\ from this table. It is also
apparent that there exist regularities in these values, which
we have studied in this work.  In particular, the strong dependence on
the true value of $\deltacp$ has turned out to be rather complicated,
though qualitatively understandable on a quite technical level. It
changes from a low-statistics (or systematics) dominated scheme to a
$\sin \deltacp$-term dominated scheme (from the oscillation
probabilities) as function of $\stheta$.  In addition, there is in
many cases a strong non-Gaussian dependence on the confidence level.
For example, for $\stheta=0.001$, $\deltacp=\pi/2$, the CP coverage
for \NuFactII\ changes more than a factor of six from the $1
\sigma$ to $3 \sigma$ confidence level. This non-Gaussian behavior can
be understood in terms of degeneracies, which are usually not present
below a certain confidence level. As an important consequence, we
conclude that it does not make sense to compare the CP coverage of
different experiments for selected parameter values, because one can
come to almost any conclusion. Only a complete comparison as function of the
full range of simulated values of $\stheta$ and $\deltacp$ can support
such conclusions, where the comparisons should be performed at rather
high confidence levels (\eg, $3 \sigma$) to include the effects of the
degeneracies.

% Problems with neutrino factory parameter space:

We have discussed the very complicated topology of the neutrino factory
parameter space which requires the highest level of sophistication.
The following major complications for the analysis and interpretation
of $\deltacp$ precision measurements at neutrino factories were
encountered: \bi
\item The topology becomes rather flat for small values of $\stheta$,
  leading to the presence of many local minima, \ie, the degeneracies
  are often present in complicated configurations.
\item The final CP coverage depends on the relative position of the
  best-fit solution and $\mathrm{sgn}(\ldm)$-degeneracy. In
  particular, for the true $\deltacp$ in the third and fourth
  quadrants, the $\mathrm{sgn}(\ldm)$-degeneracy moves as function of
  the simulated parameter values in the $\deltacp$ fit space. This
  leads to complicated parameter dependencies.
\item Depending on the parameter values, each of the best-fit solution
  and $\mathrm{sgn}(\ldm)$-degeneracy may have a
  $(\deltacp,\theta_{13})$ clone, which could actually double the CP
  coverage. Thus, in total, the CP coverage can vary by about a factor
  of four only by changing the simulated value of $\deltacp$ and thus
  the topology of the degeneracies.  Here also the strong non-Gaussian
  dependence on the confidence level originates.
\item For large values of $\stheta$, neutrino factories can be highly
  affected by matter density uncertainties.
\item Correlations {\em other} than with $\stheta$ can not be neglected.
  In some cases, their effect can even be as large as $100\%$ (\cf,
  \figu{cpcoverage}).
\item Neutrino factories contain good spectral information, which means
  that they can not be easily understood on the oscillation
  probability level.  \ei

\begin{figure}[t!]
\begin{center}
  \includegraphics[width=11cm]{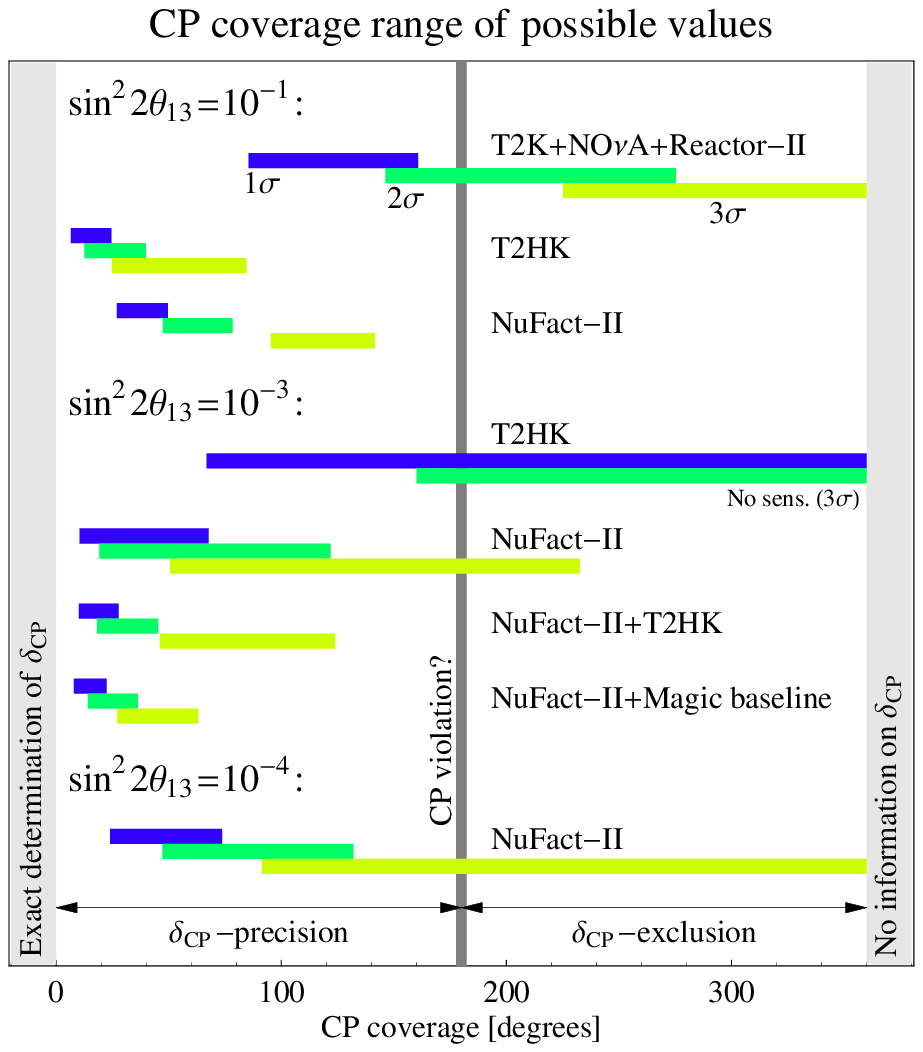}
\end{center}
\mycaption{\label{fig:covsummary} The ranges of possible values of the
  CP coverage for selected true values of $\stheta$ and experiments as
  given in the figure.
 The possible values of the CP coverage are
  given by the bands at the $1\sigma$, $2\sigma$, and $3 \sigma$
  confidence levels (from left to right). The bands therefore
  represent the possible outcomes of a CP precision measurements
  depending on the actual value of $\deltacp$, \ie, they are an
  indicator for risk minimization of a CP precision measurement.
  The left half of the figure corresponds to CP precision measurements,
  and the right half to CP exclusion measurements, where the vertical
  line in the middle is a very crude estimate for CP violation
  capabilities (since the bands are not shown for a specific value of
  $\deltacp$, such as maximal CP violation $\deltacp = \pi/2$).  }
\end{figure}

% Summary plot: That is what finally comes out there:

We summarize the impact of the true value of $\deltacp$ at different
confidence levels for different selected setups and simulated values
of $\stheta$ in \figu{covsummary}. In this figure, the bands reflect
the possible values of the CP coverage due to the unknown
true value of $\deltacp$. This means that the CP coverage at a chosen
confidence level may lay anywhere within each band.  As a matter of
high precision and risk minimization, one therefore wants to have the
{\em right} edges of the bands as far
left as possible. In addition, because
of the non-Gaussian dependence on the confidence level in many cases,
one wants to have good performances at the $3 \sigma$ confidence
level.

% Synergies and conclusion of future strategy

From the discussion in this study,
one can finally derive a comprehensive, clean strategy to $\deltacp$
precision measurements with the discussed experiments. For this purpose,
we have compared the dependence on the true values of $\stheta$ and
$\deltacp$ within the full relevant ranges. The comparison as function
of the true value of $\stheta$ has
illustrated that different experiment classes clearly operate in
different $\stheta$ regimes. The discussion as function of the true
value of $\deltacp$ has shown that only choosing specific values does
not reflect an objective comparison. For example, choosing
$\deltacp=0$, $\pi/2$, $\pi$, and $3 \pi/2$ is misleading for a
neutrino factory, because its behavior is not extreme at these values
(but somewhere in between). We finally conclude for $\deltacp$:
\begin{description}
\item[$\boldsymbol{\stheta \gtrsim 10^{-2}}$:] In this case,
  $\stheta>0$ will be established by a reactor experiment or
  superbeam. As far as $\deltacp$ precision is concerned, the choice
  to go for is a superbeam upgrade, such as \JHFHK . A neutrino factory
  is not needed in this region for a precision determination of $\deltacp$.
\item[$\boldsymbol{\stheta \lesssim 10^{-2}}$:] For $\deltacp$
  precision, one wants to have a neutrino factory at a baseline of $3
  \, 000 \, \mathrm{km}$, but with the option to add another section
  to the muon storage ring and a detector
  at a baseline of $\sim 7 \, 500 \, \mathrm{km}$ later (staged neutrino
  factory approach).  This approach
  allows to first look for $\stheta$ down to $\sim 10^{-3}$, and check
  the crude value of $\deltacp$ once $\stheta>0$ is established. If
  $\stheta$ turns out to be smaller than $\sim 10^{-3}$ or $\deltacp$
  turns out to be in the third or fourth quadrants, the operation of
  the second ``magic'' baseline becomes necessary.
\end{description}
Note that we have not included a higher gamma $\beta$-beam in this
discussion, which could, if technically feasible and competitive,
replace the role of the neutrino factory. Furthermore, a substantial
deviation from maximal atmospheric mixing or different physics
requirements (such as unitarity tests) could require additional
experiments. However, the combination between a neutrino factory and a
superbeam upgrade could not be used if $\stheta$ turned out to be
smaller than $10^{-3}$. In addition, if one assumes at least one
neutrino factory baseline anyway, one will from the point of view of
neutrino physics have to weigh the effort of
a megaton-size water Cherenkov detector (plus another accelerator
facility) against the large decay tunnel slope for a second neutrino
factory baseline.  Therefore, one should allow for a second ``magic'' baseline
from the beginning in the planning of a neutrino factory. This may actually not
 mean that one has to operate or dig the steep decay tunnel already from the beginning -- one
could only plan the option to do so later.

\subsubsection*{Acknowledgments}

We would like to thank Steve Geer for useful discussions, which
actually motivated this study.
This work has been supported by SFB 375 of Deutsche
Forschungsgemeinschaft and the W. M. Keck Foundation [WW].

%%%%%%%%%%%%%%%%%%%%%%%%%%%%%%%%%%%%%%%%%%%%%%%%%%%%%%%%%%%%%%%%%%%%%%
%%%%%%%%%%             References                         %%%%%%%%%%%%
%%%%%%%%%%%%%%%%%%%%%%%%%%%%%%%%%%%%%%%%%%%%%%%%%%%%%%%%%%%%%%%%%%%%%%

%\newpage 


\begin{thebibliography}{100}
\expandafter\ifx\csname bibnamefont\endcsname\relax
  \def\bibnamefont#1{#1}\fi
\expandafter\ifx\csname bibfnamefont\endcsname\relax
  \def\bibfnamefont#1{#1}\fi
\expandafter\ifx\csname url\endcsname\relax
  \def\url#1{\texttt{#1}}\fi
\expandafter\ifx\csname urlprefix\endcsname\relax\def\urlprefix{URL }\fi
\providecommand{\bibinfo}[2]{#2}
\providecommand{\eprint}[2][]{\url{#2}}

\bibitem{Maltoni:2004ei}
\bibinfo{author}{\bibfnamefont{M.}~\bibnamefont{Maltoni}},
  \bibinfo{author}{\bibfnamefont{T.}~\bibnamefont{Schwetz}},
  \bibinfo{author}{\bibfnamefont{M.~A.} \bibnamefont{Tortola}},
  \bibnamefont{and} \bibinfo{author}{\bibfnamefont{J.~W.~F.}
  \bibnamefont{Valle}}  (\bibinfo{year}{2004}), \eprint{hep-ph/0405172}.

\bibitem{Anderson:2004pk}
\bibinfo{author}{\bibfnamefont{K.}~\bibnamefont{Anderson}} \emph{et~al.}
  (\bibinfo{year}{2004}), \eprint{hep-ex/0402041}.

\bibitem{Ables:1995wq}
\bibinfo{author}{\bibfnamefont{E.}~\bibnamefont{Ables}} \emph{et~al.}
  (\bibinfo{collaboration}{MINOS}) \bibinfo{note}{FERMILAB-PROPOSAL-P-875}.

\bibitem{Duchesneau:2002yq}
\bibinfo{author}{\bibfnamefont{D.}~\bibnamefont{Duchesneau}}
  (\bibinfo{collaboration}{{OPERA}}), \bibinfo{journal}{eConf}
  \textbf{\bibinfo{volume}{C0209101}}, \bibinfo{pages}{TH09}
  (\bibinfo{year}{2002}), \eprint{hep-ex/0209082}.

\bibitem{Aprili:2002wx}
\bibinfo{author}{\bibfnamefont{P.}~\bibnamefont{Aprili}} \emph{et~al.}
  (\bibinfo{collaboration}{ICARUS}) \bibinfo{note}{CERN-SPSC-2002-027}.

\bibitem{Itow:2001ee}
\bibinfo{author}{\bibfnamefont{Y.}~\bibnamefont{Itow}} \emph{et~al.},
  \bibinfo{journal}{Nucl. Phys. Proc. Suppl.} \textbf{\bibinfo{volume}{111}},
  \bibinfo{pages}{146} (\bibinfo{year}{2001}),
  \eprint[http://arXiv.org/abs]{hep-ex/0106019}.

\bibitem{Ayres:2002nm}
\bibinfo{author}{\bibfnamefont{D.}~\bibnamefont{Ayres}} \emph{et~al.}
  (\bibinfo{year}{2002}), \eprint[http://arXiv.org/abs]{hep-ex/0210005}.

\bibitem{Zucchelli:2002sa}
\bibinfo{author}{\bibfnamefont{P.}~\bibnamefont{Zucchelli}},
  \bibinfo{journal}{Phys. Lett.} \textbf{\bibinfo{volume}{B532}},
  \bibinfo{pages}{166} (\bibinfo{year}{2002}).

\bibitem{Burguet-Castell:2003vv}
\bibinfo{author}{\bibfnamefont{J.}~\bibnamefont{Burguet-Castell}},
  \bibinfo{author}{\bibfnamefont{D.}~\bibnamefont{Casper}},
  \bibinfo{author}{\bibfnamefont{J.~J.} \bibnamefont{Gomez-Cadenas}},
  \bibinfo{author}{\bibfnamefont{P.}~\bibnamefont{Hernandez}},
  \bibnamefont{and} \bibinfo{author}{\bibfnamefont{F.}~\bibnamefont{Sanchez}},
  \bibinfo{journal}{Nucl. Phys.} \textbf{\bibinfo{volume}{B695}},
  \bibinfo{pages}{217} (\bibinfo{year}{2004}), \eprint{hep-ph/0312068}.

\bibitem{Geer:1998iz}
\bibinfo{author}{\bibfnamefont{S.}~\bibnamefont{Geer}}, \bibinfo{journal}{Phys.
  Rev.} \textbf{\bibinfo{volume}{D57}}, \bibinfo{pages}{6989}
  (\bibinfo{year}{1998}), \eprint[http://arXiv.org/abs]{hep-ph/9712290}.

\bibitem{Apollonio:2002en}
\bibinfo{author}{\bibfnamefont{M.}~\bibnamefont{Apollonio}} \emph{et~al.}
  (\bibinfo{year}{2002}), \eprint[http://arXiv.org/abs]{hep-ph/0210192}.

\bibitem{Albright:2004iw}
\bibinfo{author}{\bibfnamefont{C.}~\bibnamefont{Albright}} \emph{et~al.}
  (\bibinfo{collaboration}{Neutrino Factory/Muon Collider})
  (\bibinfo{year}{2004}), \eprint{physics/0411123}.

\bibitem{Barger:1980jm}
\bibinfo{author}{\bibfnamefont{V.~D.} \bibnamefont{Barger}},
  \bibinfo{author}{\bibfnamefont{K.}~\bibnamefont{Whisnant}}, \bibnamefont{and}
  \bibinfo{author}{\bibfnamefont{R.~J.~N.} \bibnamefont{Phillips}},
  \bibinfo{journal}{Phys. Rev. Lett.} \textbf{\bibinfo{volume}{45}},
  \bibinfo{pages}{2084} (\bibinfo{year}{1980}).

\bibitem{Arafune:1996bt}
\bibinfo{author}{\bibfnamefont{J.}~\bibnamefont{Arafune}} \bibnamefont{and}
  \bibinfo{author}{\bibfnamefont{J.}~\bibnamefont{Sato}},
  \bibinfo{journal}{Phys. Rev.} \textbf{\bibinfo{volume}{D55}},
  \bibinfo{pages}{1653} (\bibinfo{year}{1997}), \eprint{hep-ph/9607437}.

\bibitem{Tanimoto:1996by}
\bibinfo{author}{\bibfnamefont{M.}~\bibnamefont{Tanimoto}},
  \bibinfo{journal}{Prog. Theor. Phys.} \textbf{\bibinfo{volume}{97}},
  \bibinfo{pages}{901} (\bibinfo{year}{1997}), \eprint{hep-ph/9612444}.

\bibitem{Tanimoto:1996ky}
\bibinfo{author}{\bibfnamefont{M.}~\bibnamefont{Tanimoto}},
  \bibinfo{journal}{Phys. Rev.} \textbf{\bibinfo{volume}{D55}},
  \bibinfo{pages}{322} (\bibinfo{year}{1997}), \eprint{hep-ph/9605413}.

\bibitem{Arafune:1997hd}
\bibinfo{author}{\bibfnamefont{J.}~\bibnamefont{Arafune}},
  \bibinfo{author}{\bibfnamefont{M.}~\bibnamefont{Koike}}, \bibnamefont{and}
  \bibinfo{author}{\bibfnamefont{J.}~\bibnamefont{Sato}},
  \bibinfo{journal}{Phys. Rev.} \textbf{\bibinfo{volume}{D56}},
  \bibinfo{pages}{3093} (\bibinfo{year}{1997}), \eprint{hep-ph/9703351}.

\bibitem{Bilenky:1997dd}
\bibinfo{author}{\bibfnamefont{S.~M.} \bibnamefont{Bilenky}},
  \bibinfo{author}{\bibfnamefont{C.}~\bibnamefont{Giunti}}, \bibnamefont{and}
  \bibinfo{author}{\bibfnamefont{W.}~\bibnamefont{Grimus}},
  \bibinfo{journal}{Phys. Rev.} \textbf{\bibinfo{volume}{D58}},
  \bibinfo{pages}{033001} (\bibinfo{year}{1998}), \eprint{hep-ph/9712537}.

\bibitem{Koike:1997dh}
\bibinfo{author}{\bibfnamefont{M.}~\bibnamefont{Koike}} \bibnamefont{and}
  \bibinfo{author}{\bibfnamefont{J.}~\bibnamefont{Sato}}
  (\bibinfo{year}{1997}), \eprint{hep-ph/9707203}.

\bibitem{Minakata:1997td}
\bibinfo{author}{\bibfnamefont{H.}~\bibnamefont{Minakata}} \bibnamefont{and}
  \bibinfo{author}{\bibfnamefont{H.}~\bibnamefont{Nunokawa}},
  \bibinfo{journal}{Phys. Lett.} \textbf{\bibinfo{volume}{B413}},
  \bibinfo{pages}{369} (\bibinfo{year}{1997}), \eprint{hep-ph/9706281}.

\bibitem{Minakata:1998bf}
\bibinfo{author}{\bibfnamefont{H.}~\bibnamefont{Minakata}} \bibnamefont{and}
  \bibinfo{author}{\bibfnamefont{H.}~\bibnamefont{Nunokawa}},
  \bibinfo{journal}{Phys. Rev.} \textbf{\bibinfo{volume}{D57}},
  \bibinfo{pages}{4403} (\bibinfo{year}{1998}), \eprint{hep-ph/9705208}.

\bibitem{Tanimoto:1998sn}
\bibinfo{author}{\bibfnamefont{M.}~\bibnamefont{Tanimoto}},
  \bibinfo{journal}{Phys. Lett.} \textbf{\bibinfo{volume}{B435}},
  \bibinfo{pages}{373} (\bibinfo{year}{1998}), \eprint{hep-ph/9806375}.

\bibitem{DeRujula:1998hd}
\bibinfo{author}{\bibfnamefont{A.}~\bibnamefont{De~Rujula}},
  \bibinfo{author}{\bibfnamefont{M.~B.} \bibnamefont{Gavela}},
  \bibnamefont{and}
  \bibinfo{author}{\bibfnamefont{P.}~\bibnamefont{Hernandez}},
  \bibinfo{journal}{Nucl. Phys.} \textbf{\bibinfo{volume}{B547}},
  \bibinfo{pages}{21} (\bibinfo{year}{1999}), \eprint{hep-ph/9811390}.

\bibitem{Barger:1999fs}
\bibinfo{author}{\bibfnamefont{V.}~\bibnamefont{Barger}},
  \bibinfo{author}{\bibfnamefont{S.}~\bibnamefont{Geer}}, \bibnamefont{and}
  \bibinfo{author}{\bibfnamefont{K.}~\bibnamefont{Whisnant}},
  \bibinfo{journal}{Phys. Rev.} \textbf{\bibinfo{volume}{D61}},
  \bibinfo{pages}{053004} (\bibinfo{year}{2000}), \eprint{\hfill \\
  hep-ph/9906487}.

\bibitem{Dick:1999ed}
\bibinfo{author}{\bibfnamefont{K.}~\bibnamefont{Dick}},
  \bibinfo{author}{\bibfnamefont{M.}~\bibnamefont{Freund}},
  \bibinfo{author}{\bibfnamefont{M.}~\bibnamefont{Lindner}}, \bibnamefont{and}
  \bibinfo{author}{\bibfnamefont{A.}~\bibnamefont{Romanino}},
  \bibinfo{journal}{Nucl. Phys.} \textbf{\bibinfo{volume}{B562}},
  \bibinfo{pages}{29} (\bibinfo{year}{1999}), \eprint{hep-ph/9903308}.

\bibitem{Donini:1999jc}
\bibinfo{author}{\bibfnamefont{A.}~\bibnamefont{Donini}},
  \bibinfo{author}{\bibfnamefont{M.~B.} \bibnamefont{Gavela}},
  \bibinfo{author}{\bibfnamefont{P.}~\bibnamefont{Hernandez}},
  \bibnamefont{and} \bibinfo{author}{\bibfnamefont{S.}~\bibnamefont{Rigolin}},
  \bibinfo{journal}{Nucl. Phys.} \textbf{\bibinfo{volume}{B574}},
  \bibinfo{pages}{23} (\bibinfo{year}{2000}), \eprint{hep-ph/9909254}.

\bibitem{Freund:1999gy}
\bibinfo{author}{\bibfnamefont{M.}~\bibnamefont{Freund}},
  \bibinfo{author}{\bibfnamefont{M.}~\bibnamefont{Lindner}},
  \bibinfo{author}{\bibfnamefont{S.~T.} \bibnamefont{Petcov}},
  \bibnamefont{and} \bibinfo{author}{\bibfnamefont{A.}~\bibnamefont{Romanino}},
  \bibinfo{journal}{Nucl. Phys.} \textbf{\bibinfo{volume}{B578}},
  \bibinfo{pages}{27} (\bibinfo{year}{2000}), \eprint{hep-ph/9912457}.

\bibitem{Koike:1999hf}
\bibinfo{author}{\bibfnamefont{M.}~\bibnamefont{Koike}} \bibnamefont{and}
  \bibinfo{author}{\bibfnamefont{J.}~\bibnamefont{Sato}},
  \bibinfo{journal}{Phys. Rev.} \textbf{\bibinfo{volume}{D61}},
  \bibinfo{pages}{073012} (\bibinfo{year}{2000}), \eprint{hep-ph/9909469}.

\bibitem{Romanino:1999zq}
\bibinfo{author}{\bibfnamefont{A.}~\bibnamefont{Romanino}},
  \bibinfo{journal}{Nucl. Phys.} \textbf{\bibinfo{volume}{B574}},
  \bibinfo{pages}{675} (\bibinfo{year}{2000}),
  \eprint[http://arXiv.org/abs]{hep-ph/9909425}.

\bibitem{Sato:1999wt}
\bibinfo{author}{\bibfnamefont{J.}~\bibnamefont{Sato}}, \bibinfo{journal}{Nucl.
  Instrum. Meth.} \textbf{\bibinfo{volume}{A451}}, \bibinfo{pages}{36}
  (\bibinfo{year}{2000}), \eprint{hep-ph/9910442}.

\bibitem{Cervera:2000kp}
\bibinfo{author}{\bibfnamefont{A.}~\bibnamefont{Cervera}} \emph{et~al.},
  \bibinfo{journal}{Nucl. Phys.} \textbf{\bibinfo{volume}{B579}},
  \bibinfo{pages}{17} (\bibinfo{year}{2000}), \eprint{hep-ph/0002108}.

\bibitem{Minakata:2000fe}
\bibinfo{author}{\bibfnamefont{H.}~\bibnamefont{Minakata}} \bibnamefont{and}
  \bibinfo{author}{\bibfnamefont{H.}~\bibnamefont{Nunokawa}},
  \bibinfo{journal}{Nucl. Instrum. Meth.} \textbf{\bibinfo{volume}{A472}},
  \bibinfo{pages}{421} (\bibinfo{year}{2000}), \eprint{hep-ph/0009091}.

\bibitem{Minakata:2000ee}
\bibinfo{author}{\bibfnamefont{H.}~\bibnamefont{Minakata}} \bibnamefont{and}
  \bibinfo{author}{\bibfnamefont{H.}~\bibnamefont{Nunokawa}},
  \bibinfo{journal}{Phys. Lett.} \textbf{\bibinfo{volume}{B495}},
  \bibinfo{pages}{369} (\bibinfo{year}{2000}), \eprint{hep-ph/0004114}.

\bibitem{Koike:2000jf}
\bibinfo{author}{\bibfnamefont{M.}~\bibnamefont{Koike}},
  \bibinfo{author}{\bibfnamefont{T.}~\bibnamefont{Ota}}, \bibnamefont{and}
  \bibinfo{author}{\bibfnamefont{J.}~\bibnamefont{Sato}},
  \bibinfo{journal}{Nucl. Phys.} \textbf{\bibinfo{volume}{B615}},
  \bibinfo{pages}{331}, \eprint{hep-ph/0011387}.

\bibitem{Barger:2001yr}
\bibinfo{author}{\bibfnamefont{V.}~\bibnamefont{Barger}},
  \bibinfo{author}{\bibfnamefont{D.}~\bibnamefont{Marfatia}}, \bibnamefont{and}
  \bibinfo{author}{\bibfnamefont{K.}~\bibnamefont{Whisnant}},
  \bibinfo{journal}{Phys. Rev.} \textbf{\bibinfo{volume}{D65}},
  \bibinfo{pages}{073023} (\bibinfo{year}{2002}),
  \eprint[http://arXiv.org/abs]{hep-ph/0112119}.

\bibitem{Bueno:2001jd}
\bibinfo{author}{\bibfnamefont{A.}~\bibnamefont{Bueno}},
  \bibinfo{author}{\bibfnamefont{M.}~\bibnamefont{Campanelli}},
  \bibinfo{author}{\bibfnamefont{S.}~\bibnamefont{Navas-Concha}},
  \bibnamefont{and} \bibinfo{author}{\bibfnamefont{A.}~\bibnamefont{Rubbia}},
  \bibinfo{journal}{Nucl. Phys.} \textbf{\bibinfo{volume}{B631}},
  \bibinfo{pages}{239} (\bibinfo{year}{2002}),
  \eprint[http://arXiv.org/abs]{hep-ph/0112297}.

\bibitem{Burguet-Castell:2001ez}
\bibinfo{author}{\bibfnamefont{J.}~\bibnamefont{Burguet-Castell}},
  \bibinfo{author}{\bibfnamefont{M.~B.} \bibnamefont{Gavela}},
  \bibinfo{author}{\bibfnamefont{J.~J.} \bibnamefont{Gomez-Cadenas}},
  \bibinfo{author}{\bibfnamefont{P.}~\bibnamefont{Hernandez}},
  \bibnamefont{and} \bibinfo{author}{\bibfnamefont{O.}~\bibnamefont{Mena}},
  \bibinfo{journal}{Nucl. Phys.} \textbf{\bibinfo{volume}{B608}},
  \bibinfo{pages}{301} (\bibinfo{year}{2001}),
  \eprint[http://arXiv.org/abs]{hep-ph/0103258}.

\bibitem{Koike:2001kv}
\bibinfo{author}{\bibfnamefont{M.}~\bibnamefont{Koike}},
  \bibinfo{author}{\bibfnamefont{T.}~\bibnamefont{Ota}}, \bibnamefont{and}
  \bibinfo{author}{\bibfnamefont{J.}~\bibnamefont{Sato}},
  \bibinfo{journal}{Phys. Rev.} \textbf{\bibinfo{volume}{D65}},
  \bibinfo{pages}{053015}, \eprint{hep-ph/0103024}.

\bibitem{Minakata:2001rj}
\bibinfo{author}{\bibfnamefont{H.}~\bibnamefont{Minakata}} \bibnamefont{and}
  \bibinfo{author}{\bibfnamefont{H.}~\bibnamefont{Nunokawa}},
  \bibinfo{journal}{Nucl. Instrum. Meth.} \textbf{\bibinfo{volume}{A503}},
  \bibinfo{pages}{218} (\bibinfo{year}{2001}), \eprint{hep-ph/0111130}.

\bibitem{Pinney:2001xw}
\bibinfo{author}{\bibfnamefont{J.}~\bibnamefont{Pinney}} \bibnamefont{and}
  \bibinfo{author}{\bibfnamefont{O.}~\bibnamefont{Yasuda}},
  \bibinfo{journal}{Phys. Rev.} \textbf{\bibinfo{volume}{D64}},
  \bibinfo{pages}{093008} (\bibinfo{year}{2001}), \eprint{hep-ph/0105087}.

\bibitem{Aoki:2002ae}
\bibinfo{author}{\bibfnamefont{M.}~\bibnamefont{Aoki}},
  \bibinfo{author}{\bibfnamefont{K.}~\bibnamefont{Hagiwara}}, \bibnamefont{and}
  \bibinfo{author}{\bibfnamefont{N.}~\bibnamefont{Okamura}},
  \bibinfo{journal}{Phys. Lett.} \textbf{\bibinfo{volume}{B554}},
  \bibinfo{pages}{121} (\bibinfo{year}{2003}), \eprint{hep-ph/0208223}.

\bibitem{Barger:2002xk}
\bibinfo{author}{\bibfnamefont{V.}~\bibnamefont{Barger}},
  \bibinfo{author}{\bibfnamefont{D.}~\bibnamefont{Marfatia}}, \bibnamefont{and}
  \bibinfo{author}{\bibfnamefont{K.}~\bibnamefont{Whisnant}},
  \bibinfo{journal}{Phys. Lett.} \textbf{\bibinfo{volume}{B560}},
  \bibinfo{pages}{75} (\bibinfo{year}{2003}), \eprint{hep-ph/0210428}.

\bibitem{Burguet-Castell:2002qx}
\bibinfo{author}{\bibfnamefont{J.}~\bibnamefont{Burguet-Castell}},
  \bibinfo{author}{\bibfnamefont{M.~B.} \bibnamefont{Gavela}},
  \bibinfo{author}{\bibfnamefont{J.~J.} \bibnamefont{Gomez-Cadenas}},
  \bibinfo{author}{\bibfnamefont{P.}~\bibnamefont{Hernandez}},
  \bibnamefont{and} \bibinfo{author}{\bibfnamefont{O.}~\bibnamefont{Mena}},
  \bibinfo{journal}{Nucl. Phys.} \textbf{\bibinfo{volume}{B646}},
  \bibinfo{pages}{301} (\bibinfo{year}{2002}),
  \eprint[http://arXiv.org/abs]{hep-ph/0207080}.

\bibitem{Donini:2002rm}
\bibinfo{author}{\bibfnamefont{A.}~\bibnamefont{Donini}},
  \bibinfo{author}{\bibfnamefont{D.}~\bibnamefont{Meloni}}, \bibnamefont{and}
  \bibinfo{author}{\bibfnamefont{P.}~\bibnamefont{Migliozzi}},
  \bibinfo{journal}{Nucl. Phys.} \textbf{\bibinfo{volume}{B646}},
  \bibinfo{pages}{321} (\bibinfo{year}{2002}),
  \eprint[http://arXiv.org/abs]{hep-ph/0206034}.

\bibitem{Huber:2002mx}
\bibinfo{author}{\bibfnamefont{P.}~\bibnamefont{Huber}},
  \bibinfo{author}{\bibfnamefont{M.}~\bibnamefont{Lindner}}, \bibnamefont{and}
  \bibinfo{author}{\bibfnamefont{W.}~\bibnamefont{Winter}},
  \bibinfo{journal}{Nucl. Phys.} \textbf{\bibinfo{volume}{B645}},
  \bibinfo{pages}{3} (\bibinfo{year}{2002}), \eprint{hep-ph/0204352}.

\bibitem{Huber:2002rs}
\bibinfo{author}{\bibfnamefont{P.}~\bibnamefont{Huber}},
  \bibinfo{author}{\bibfnamefont{M.}~\bibnamefont{Lindner}}, \bibnamefont{and}
  \bibinfo{author}{\bibfnamefont{W.}~\bibnamefont{Winter}},
  \bibinfo{journal}{Nucl. Phys.} \textbf{\bibinfo{volume}{B654}},
  \bibinfo{pages}{3} (\bibinfo{year}{2003}), \eprint{hep-ph/0211300}.

\bibitem{Minakata:2002qi}
\bibinfo{author}{\bibfnamefont{H.}~\bibnamefont{Minakata}},
  \bibinfo{author}{\bibfnamefont{H.}~\bibnamefont{Nunokawa}}, \bibnamefont{and}
  \bibinfo{author}{\bibfnamefont{S.}~\bibnamefont{Parke}}
  (\bibinfo{year}{2002}), \eprint[http://arXiv.org/abs]{hep-ph/0208163}.

\bibitem{Minakata:2002qe}
\bibinfo{author}{\bibfnamefont{H.}~\bibnamefont{Minakata}},
  \bibinfo{author}{\bibfnamefont{H.}~\bibnamefont{Nunokawa}}, \bibnamefont{and}
  \bibinfo{author}{\bibfnamefont{S.~J.} \bibnamefont{Parke}},
  \bibinfo{journal}{Phys. Lett.} \textbf{\bibinfo{volume}{B537}},
  \bibinfo{pages}{249} (\bibinfo{year}{2002}), \eprint{hep-ph/0204171}.

\bibitem{Ota:2002fu}
\bibinfo{author}{\bibfnamefont{T.}~\bibnamefont{Ota}} \bibnamefont{and}
  \bibinfo{author}{\bibfnamefont{J.}~\bibnamefont{Sato}}
  (\bibinfo{year}{2002}), \eprint[http://arXiv.org/abs]{hep-ph/0211095}.

\bibitem{Whisnant:2002fx}
\bibinfo{author}{\bibfnamefont{K.}~\bibnamefont{Whisnant}},
  \bibinfo{author}{\bibfnamefont{J.~M.} \bibnamefont{Yang}}, \bibnamefont{and}
  \bibinfo{author}{\bibfnamefont{B.-L.} \bibnamefont{Young}},
  \bibinfo{journal}{Phys. Rev.} \textbf{\bibinfo{volume}{D67}},
  \bibinfo{pages}{013004} (\bibinfo{year}{2003}), \eprint{hep-ph/0208193}.

\bibitem{Autiero:2003fu}
\bibinfo{author}{\bibfnamefont{D.}~\bibnamefont{Autiero}} \emph{et~al.}
  (\bibinfo{year}{2003}), \eprint{hep-ph/0305185}.

\bibitem{Brahmachari:2003bk}
\bibinfo{author}{\bibfnamefont{B.}~\bibnamefont{Brahmachari}},
  \bibinfo{author}{\bibfnamefont{S.}~\bibnamefont{Choubey}}, \bibnamefont{and}
  \bibinfo{author}{\bibfnamefont{P.}~\bibnamefont{Roy}},
  \bibinfo{journal}{Nucl. Phys.} \textbf{\bibinfo{volume}{B671}},
  \bibinfo{pages}{483} (\bibinfo{year}{2003}), \eprint{hep-ph/0303078}.

\bibitem{Diwan:2003bp}
\bibinfo{author}{\bibfnamefont{M.~V.} \bibnamefont{Diwan}} \emph{et~al.}
  (\bibinfo{year}{2003}), \eprint{hep-ph/0303081}.

\bibitem{Donini:2003vz}
\bibinfo{author}{\bibfnamefont{A.}~\bibnamefont{Donini}},
  \bibinfo{author}{\bibfnamefont{D.}~\bibnamefont{Meloni}}, \bibnamefont{and}
  \bibinfo{author}{\bibfnamefont{S.}~\bibnamefont{Rigolin}},
  \bibinfo{journal}{JHEP} \textbf{\bibinfo{volume}{06}}, \bibinfo{pages}{011}
  (\bibinfo{year}{2004}), \eprint{hep-ph/0312072}.

\bibitem{Huber:2003ak}
\bibinfo{author}{\bibfnamefont{P.}~\bibnamefont{Huber}} \bibnamefont{and}
  \bibinfo{author}{\bibfnamefont{W.}~\bibnamefont{Winter}},
  \bibinfo{journal}{Phys. Rev.} \textbf{\bibinfo{volume}{D68}},
  \bibinfo{pages}{037301} (\bibinfo{year}{2003}), \eprint{hep-ph/0301257}.

\bibitem{Mezzetto:2003ub}
\bibinfo{author}{\bibfnamefont{M.}~\bibnamefont{Mezzetto}},
  \bibinfo{journal}{J. Phys.} \textbf{\bibinfo{volume}{G29}},
  \bibinfo{pages}{1771} (\bibinfo{year}{2003}), \eprint{hep-ex/0302007}.

\bibitem{Migliozzi:2003pw}
\bibinfo{author}{\bibfnamefont{P.}~\bibnamefont{Migliozzi}} \bibnamefont{and}
  \bibinfo{author}{\bibfnamefont{F.}~\bibnamefont{Terranova}},
  \bibinfo{journal}{Phys. Lett.} \textbf{\bibinfo{volume}{B563}},
  \bibinfo{pages}{73} (\bibinfo{year}{2003}), \eprint{hep-ph/0302274}.

\bibitem{Minakata:2003wq}
\bibinfo{author}{\bibfnamefont{H.}~\bibnamefont{Minakata}} \bibnamefont{and}
  \bibinfo{author}{\bibfnamefont{H.}~\bibnamefont{Sugiyama}},
  \bibinfo{journal}{Phys. Lett.} \textbf{\bibinfo{volume}{B580}},
  \bibinfo{pages}{216} (\bibinfo{year}{2004}), \eprint{hep-ph/0309323}.

\bibitem{Ohlsson:2003ip}
\bibinfo{author}{\bibfnamefont{T.}~\bibnamefont{Ohlsson}} \bibnamefont{and}
  \bibinfo{author}{\bibfnamefont{W.}~\bibnamefont{Winter}},
  \bibinfo{journal}{Phys. Rev.} \textbf{\bibinfo{volume}{D68}},
  \bibinfo{pages}{073007} (\bibinfo{year}{2003}), \eprint{hep-ph/0307178}.

\bibitem{Shan:2003vh}
\bibinfo{author}{\bibfnamefont{L.-Y.} \bibnamefont{Shan}} \emph{et~al.},
  \bibinfo{journal}{Phys. Rev.} \textbf{\bibinfo{volume}{D68}},
  \bibinfo{pages}{013002} (\bibinfo{year}{2003}), \eprint{hep-ph/0303112}.

\bibitem{Winter:2003ye}
\bibinfo{author}{\bibfnamefont{W.}~\bibnamefont{Winter}},
  \bibinfo{journal}{Phys. Rev.} \textbf{\bibinfo{volume}{D70}},
  \bibinfo{pages}{033006} (\bibinfo{year}{2004}), \eprint{hep-ph/0310307}.

\bibitem{Blom:2004bk}
\bibinfo{author}{\bibfnamefont{M.}~\bibnamefont{Blom}} \bibnamefont{and}
  \bibinfo{author}{\bibfnamefont{H.}~\bibnamefont{Minakata}},
  \bibinfo{journal}{New J. Phys.} \textbf{\bibinfo{volume}{6}},
  \bibinfo{pages}{130} (\bibinfo{year}{2004}), \eprint{hep-ph/0404142}.

\bibitem{Huber:2004ug}
\bibinfo{author}{\bibfnamefont{P.}~\bibnamefont{Huber}},
  \bibinfo{author}{\bibfnamefont{M.}~\bibnamefont{Lindner}},
  \bibinfo{author}{\bibfnamefont{M.}~\bibnamefont{Rolinec}},
  \bibinfo{author}{\bibfnamefont{T.}~\bibnamefont{Schwetz}}, \bibnamefont{and}
  \bibinfo{author}{\bibfnamefont{W.}~\bibnamefont{Winter}},
  \bibinfo{journal}{Phys. Rev.} \textbf{\bibinfo{volume}{D70}},
  \bibinfo{pages}{073014} (\bibinfo{year}{2004}), \eprint{hep-ph/0403068}.

\bibitem{Mena:2004sa}
\bibinfo{author}{\bibfnamefont{O.}~\bibnamefont{Mena}} \bibnamefont{and}
  \bibinfo{author}{\bibfnamefont{S.}~\bibnamefont{Parke}},
  \bibinfo{journal}{Phys. Rev.} \textbf{\bibinfo{volume}{D70}},
  \bibinfo{pages}{093011} (\bibinfo{year}{2004}), \eprint{hep-ph/0408070}.

\bibitem{Minakata:2004vz}
\bibinfo{author}{\bibfnamefont{H.}~\bibnamefont{Minakata}}
  (\bibinfo{year}{2004}), \eprint{hep-ph/0402197}.

\bibitem{Ambats:2004js}
\bibinfo{author}{\bibfnamefont{I.}~\bibnamefont{Ambats}} \emph{et~al.}
  (\bibinfo{collaboration}{NOvA}) \bibinfo{note}{FERMILAB-PROPOSAL-0929}.

\bibitem{Freund:2001pn}
\bibinfo{author}{\bibfnamefont{M.}~\bibnamefont{Freund}},
  \bibinfo{journal}{Phys. Rev.} \textbf{\bibinfo{volume}{D64}},
  \bibinfo{pages}{053003} (\bibinfo{year}{2001}), \eprint{hep-ph/0103300}.

\bibitem{Akhmedov:2004ny}
\bibinfo{author}{\bibfnamefont{E.~K.} \bibnamefont{Akhmedov}},
  \bibinfo{author}{\bibfnamefont{R.}~\bibnamefont{Johansson}},
  \bibinfo{author}{\bibfnamefont{M.}~\bibnamefont{Lindner}},
  \bibinfo{author}{\bibfnamefont{T.}~\bibnamefont{Ohlsson}}, \bibnamefont{and}
  \bibinfo{author}{\bibfnamefont{T.}~\bibnamefont{Schwetz}}
  (\bibinfo{year}{2004}), \eprint{hep-ph/0402175}.

\bibitem{Bahcall:2004ut}
\bibinfo{author}{\bibfnamefont{J.~N.} \bibnamefont{Bahcall}},
  \bibinfo{author}{\bibfnamefont{M.~C.} \bibnamefont{Gonzalez-Garcia}},
  \bibnamefont{and}
  \bibinfo{author}{\bibfnamefont{C.}~\bibnamefont{Pena-Garay}},
  \bibinfo{journal}{JHEP} \textbf{\bibinfo{volume}{08}}, \bibinfo{pages}{016}
  (\bibinfo{year}{2004}), \eprint{hep-ph/0406294}.

\bibitem{Bandyopadhyay:2004da}
\bibinfo{author}{\bibfnamefont{A.}~\bibnamefont{Bandyopadhyay}},
  \bibinfo{author}{\bibfnamefont{S.}~\bibnamefont{Choubey}},
  \bibinfo{author}{\bibfnamefont{S.}~\bibnamefont{Goswami}},
  \bibinfo{author}{\bibfnamefont{S.~T.} \bibnamefont{Petcov}},
  \bibnamefont{and} \bibinfo{author}{\bibfnamefont{D.~P.} \bibnamefont{Roy}}
  (\bibinfo{year}{2004}), \eprint{hep-ph/0406328}.

\bibitem{Maltoni:2003da}
\bibinfo{author}{\bibfnamefont{M.}~\bibnamefont{Maltoni}},
  \bibinfo{author}{\bibfnamefont{T.}~\bibnamefont{Schwetz}},
  \bibinfo{author}{\bibfnamefont{M.~A.} \bibnamefont{Tortola}},
  \bibnamefont{and} \bibinfo{author}{\bibfnamefont{J.~W.~F.}
  \bibnamefont{Valle}}, \bibinfo{journal}{Phys. Rev.}
  \textbf{\bibinfo{volume}{D68}}, \bibinfo{pages}{113010}
  (\bibinfo{year}{2003}), \eprint{hep-ph/0309130}.

\bibitem{Minakata:2001qm}
\bibinfo{author}{\bibfnamefont{H.}~\bibnamefont{Minakata}} \bibnamefont{and}
  \bibinfo{author}{\bibfnamefont{H.}~\bibnamefont{Nunokawa}},
  \bibinfo{journal}{JHEP} \textbf{\bibinfo{volume}{10}}, \bibinfo{pages}{001}
  (\bibinfo{year}{2001}), \eprint[http://arXiv.org/abs]{hep-ph/0108085}.

\bibitem{Fogli:1996pv}
\bibinfo{author}{\bibfnamefont{G.~L.} \bibnamefont{Fogli}} \bibnamefont{and}
  \bibinfo{author}{\bibfnamefont{E.}~\bibnamefont{Lisi}},
  \bibinfo{journal}{Phys. Rev.} \textbf{\bibinfo{volume}{D54}},
  \bibinfo{pages}{3667} (\bibinfo{year}{1996}), \eprint{hep-ph/9604415}.

\bibitem{Lipari:1999wy}
\bibinfo{author}{\bibfnamefont{P.}~\bibnamefont{Lipari}},
  \bibinfo{journal}{Phys. Rev.} \textbf{\bibinfo{volume}{D61}},
  \bibinfo{pages}{113004} (\bibinfo{year}{2000}), \eprint{hep-ph/9903481}.

\bibitem{Huber:2002uy}
\bibinfo{author}{\bibfnamefont{P.}~\bibnamefont{Huber}}, \bibinfo{journal}{J.
  Phys.} \textbf{\bibinfo{volume}{G29}}, \bibinfo{pages}{1853}
  (\bibinfo{year}{2003}), \eprint{hep-ph/0210140}.

\bibitem{Fogli:2003th}
\bibinfo{author}{\bibfnamefont{G.~L.} \bibnamefont{Fogli}},
  \bibinfo{author}{\bibfnamefont{E.}~\bibnamefont{Lisi}},
  \bibinfo{author}{\bibfnamefont{A.}~\bibnamefont{Marrone}}, \bibnamefont{and}
  \bibinfo{author}{\bibfnamefont{D.}~\bibnamefont{Montanino}},
  \bibinfo{journal}{Phys. Rev.} \textbf{\bibinfo{volume}{D67}},
  \bibinfo{pages}{093006} (\bibinfo{year}{2003}), \eprint{hep-ph/0303064}.

\bibitem{Geller:2001ix}
\bibinfo{author}{\bibfnamefont{R.~J.} \bibnamefont{Geller}} \bibnamefont{and}
  \bibinfo{author}{\bibfnamefont{T.}~\bibnamefont{Hara}},
  \bibinfo{journal}{Phys. Rev. Lett.} \textbf{\bibinfo{volume}{49}},
  \bibinfo{pages}{98} (\bibinfo{year}{2001}),
  \eprint[http://arXiv.org/abs]{hep-ph/0111342}.

\bibitem{Pana}
\bibinfo{author}{\bibfnamefont{S.~V.} \bibnamefont{Panasyuk}},
  \emph{\bibinfo{title}{Rem (reference earth model) web page}}
  (\bibinfo{year}{2000}), \bibinfo{note}{{\tt
  http://cfauvcs5.harvard.edu/lana/rem/index.htm}}.

\bibitem{Huber:2004ka}
\bibinfo{author}{\bibfnamefont{P.}~\bibnamefont{Huber}},
  \bibinfo{author}{\bibfnamefont{M.}~\bibnamefont{Lindner}}, \bibnamefont{and}
  \bibinfo{author}{\bibfnamefont{W.}~\bibnamefont{Winter}}
  (\bibinfo{year}{2004}), \bibinfo{note}{{\tt
  http://www.ph.tum.de/$^\sim$globes}}, \eprint{hep-ph/0407333}.

\bibitem{Gonzalez-Garcia:2001zy}
\bibinfo{author}{\bibfnamefont{M.~C.} \bibnamefont{Gonzalez-Garcia}}
  \bibnamefont{and}
  \bibinfo{author}{\bibfnamefont{C.}~\bibnamefont{Pe$\tilde{\mathrm{n}}$a-Gara%
y}}, \bibinfo{journal}{Phys. Lett.} \textbf{\bibinfo{volume}{B527}},
  \bibinfo{pages}{199} (\bibinfo{year}{2002}),
  \eprint[http://arXiv.org/abs]{hep-ph/0111432}.

\bibitem{Barger:2000hy}
\bibinfo{author}{\bibfnamefont{V.~D.} \bibnamefont{Barger}},
  \bibinfo{author}{\bibfnamefont{D.}~\bibnamefont{Marfatia}}, \bibnamefont{and}
  \bibinfo{author}{\bibfnamefont{B.~P.} \bibnamefont{Wood}},
  \bibinfo{journal}{Phys. Lett.} \textbf{\bibinfo{volume}{B498}},
  \bibinfo{pages}{53} (\bibinfo{year}{2001}), \eprint{hep-ph/0011251}.

\bibitem{Huber:2003pm}
\bibinfo{author}{\bibfnamefont{P.}~\bibnamefont{Huber}},
  \bibinfo{author}{\bibfnamefont{M.}~\bibnamefont{Lindner}},
  \bibinfo{author}{\bibfnamefont{T.}~\bibnamefont{Schwetz}}, \bibnamefont{and}
  \bibinfo{author}{\bibfnamefont{W.}~\bibnamefont{Winter}},
  \bibinfo{journal}{Nucl. Phys.} \textbf{\bibinfo{volume}{B665}},
  \bibinfo{pages}{487} (\bibinfo{year}{2003}), \eprint{hep-ph/0303232}.

\bibitem{betainprep}
\bibinfo{author}{\bibfnamefont{P.}~\bibnamefont{Huber}},
  \bibinfo{author}{\bibfnamefont{M.}~\bibnamefont{Lindner}},
  \bibinfo{author}{\bibfnamefont{M.}~\bibnamefont{Rolinec}}, \bibnamefont{and}
  \bibinfo{author}{\bibfnamefont{W.}~\bibnamefont{Winter}}, \bibinfo{note}{in
  preparation}.

\bibitem{Winter:2003st}
\bibinfo{author}{\bibfnamefont{W.}~\bibnamefont{Winter}}, \bibinfo{journal}{AIP
  Conf. Proc.} \textbf{\bibinfo{volume}{721}}, \bibinfo{pages}{227}
  (\bibinfo{year}{2004}), \eprint{hep-ph/0308227}.

\bibitem{offaxis}
\bibinfo{author}{\bibfnamefont{D.}~\bibnamefont{Beavis}} \emph{et~al.},
  \emph{\bibinfo{title}{Proposal of BNL AGS E-889}}, \bibinfo{type}{Tech.
  Rep.}, \bibinfo{institution}{BNL} (\bibinfo{year}{1995}).

\bibitem{Minakata:2003ca}
\bibinfo{author}{\bibfnamefont{H.}~\bibnamefont{Minakata}},
  \bibinfo{author}{\bibfnamefont{H.}~\bibnamefont{Nunokawa}}, \bibnamefont{and}
  \bibinfo{author}{\bibfnamefont{S.~J.} \bibnamefont{Parke}},
  \bibinfo{journal}{Phys. Rev.} \textbf{\bibinfo{volume}{D68}},
  \bibinfo{pages}{013010} (\bibinfo{year}{2003}), \eprint{hep-ph/0301210}.

\bibitem{Minakata:2002jv}
\bibinfo{author}{\bibfnamefont{H.}~\bibnamefont{Minakata}},
  \bibinfo{author}{\bibfnamefont{H.}~\bibnamefont{Sugiyama}},
  \bibinfo{author}{\bibfnamefont{O.}~\bibnamefont{Yasuda}},
  \bibinfo{author}{\bibfnamefont{K.}~\bibnamefont{Inoue}}, \bibnamefont{and}
  \bibinfo{author}{\bibfnamefont{F.}~\bibnamefont{Suekane}},
  \bibinfo{journal}{Phys. Rev.} \textbf{\bibinfo{volume}{D68}},
  \bibinfo{pages}{033017} (\bibinfo{year}{2003}), \eprint{hep-ph/0211111}.

\bibitem{Wang:2001ys}
\bibinfo{author}{\bibfnamefont{Y.~F.} \bibnamefont{Wang}},
  \bibinfo{author}{\bibfnamefont{K.}~\bibnamefont{Whisnant}},
  \bibinfo{author}{\bibfnamefont{Z.-h.} \bibnamefont{Xiong}},
  \bibinfo{author}{\bibfnamefont{J.~M.} \bibnamefont{Yang}}, \bibnamefont{and}
  \bibinfo{author}{\bibfnamefont{B.-L.} \bibnamefont{Young}}
  (\bibinfo{collaboration}{VLBL Study Group H2B-4}), \bibinfo{journal}{Phys.
  Rev.} \textbf{\bibinfo{volume}{D65}}, \bibinfo{pages}{073021}
  (\bibinfo{year}{2002}), \eprint{hep-ph/0111317}.

\bibitem{Barger:2002rr}
\bibinfo{author}{\bibfnamefont{V.}~\bibnamefont{Barger}},
  \bibinfo{author}{\bibfnamefont{D.}~\bibnamefont{Marfatia}}, \bibnamefont{and}
  \bibinfo{author}{\bibfnamefont{K.}~\bibnamefont{Whisnant}},
  \bibinfo{journal}{Phys. Rev.} \textbf{\bibinfo{volume}{D66}},
  \bibinfo{pages}{053007} (\bibinfo{year}{2002}),
  \eprint[http://arXiv.org/abs]{hep-ph/0206038}.

\bibitem{Asratyan:2003dp}
\bibinfo{author}{\bibfnamefont{A.}~\bibnamefont{Asratyan}} \emph{et~al.},
  \bibinfo{journal}{Science} \textbf{\bibinfo{volume}{124}},
  \bibinfo{pages}{103} (\bibinfo{year}{2003}), \eprint{hep-ex/0303023}.

\bibitem{Donini:2004hu}
\bibinfo{author}{\bibfnamefont{A.}~\bibnamefont{Donini}},
  \bibinfo{author}{\bibfnamefont{E.}~\bibnamefont{Fernandez-Martinez}},
  \bibinfo{author}{\bibfnamefont{P.}~\bibnamefont{Migliozzi}},
  \bibinfo{author}{\bibfnamefont{S.}~\bibnamefont{Rigolin}}, \bibnamefont{and}
  \bibinfo{author}{\bibfnamefont{L.}~\bibnamefont{Scotto~Lavina}}
  (\bibinfo{year}{2004}), \eprint{hep-ph/0406132}.

\bibitem{Yasuda:2004gu}
\bibinfo{author}{\bibfnamefont{O.}~\bibnamefont{Yasuda}}
  (\bibinfo{year}{2004}), \eprint{hep-ph/0405005}.

\bibitem{Donini:2004iv}
\bibinfo{author}{\bibfnamefont{A.}~\bibnamefont{Donini}},
  \bibinfo{author}{\bibfnamefont{E.}~\bibnamefont{Fernandez-Martinez}},
  \bibnamefont{and} \bibinfo{author}{\bibfnamefont{S.}~\bibnamefont{Rigolin}}
  (\bibinfo{year}{2004}), \eprint{hep-ph/0411402}.

\bibitem{Bouchez:2003fy}
\bibinfo{author}{\bibfnamefont{J.}~\bibnamefont{Bouchez}},
  \bibinfo{author}{\bibfnamefont{M.}~\bibnamefont{Lindroos}}, \bibnamefont{and}
  \bibinfo{author}{\bibfnamefont{M.}~\bibnamefont{Mezzetto}}
  (\bibinfo{year}{2003}), \eprint{hep-ex/0310059}.

\bibitem{McConnel:2004bd}
\bibinfo{author}{\bibfnamefont{K.~B.} \bibnamefont{McConnel}} \bibnamefont{and}
  \bibinfo{author}{\bibfnamefont{M.~H.} \bibnamefont{Shaevitz}}
  (\bibinfo{year}{2004}), \eprint{hep-ex/0409028}.

\bibitem{Diwan:2004bt}
\bibinfo{author}{\bibfnamefont{M.~V.} \bibnamefont{Diwan}}
  (\bibinfo{year}{2004}), \eprint{hep-ex/0407047}.

\bibitem{Barger:2004db}
\bibinfo{author}{\bibfnamefont{V.}~\bibnamefont{Barger}},
  \bibinfo{author}{\bibfnamefont{S.}~\bibnamefont{Geer}}, \bibnamefont{and}
  \bibinfo{author}{\bibfnamefont{K.}~\bibnamefont{Whisnant}}
  (\bibinfo{year}{2004}), \eprint{hep-ph/0407140}.

\bibitem{Johnson:1999ci}
\bibinfo{author}{\bibfnamefont{L.~M.} \bibnamefont{Johnson}} \bibnamefont{and}
  \bibinfo{author}{\bibfnamefont{D.~W.} \bibnamefont{McKay}},
  \bibinfo{journal}{Phys. Rev.} \textbf{\bibinfo{volume}{D61}},
  \bibinfo{pages}{113007} (\bibinfo{year}{2000}), \eprint{hep-ph/9909355}.

\bibitem{Gago:2001xg}
\bibinfo{author}{\bibfnamefont{A.~M.} \bibnamefont{Gago}},
  \bibinfo{author}{\bibfnamefont{M.~M.} \bibnamefont{Guzzo}},
  \bibinfo{author}{\bibfnamefont{H.}~\bibnamefont{Nunokawa}},
  \bibinfo{author}{\bibfnamefont{W.~J.~C.} \bibnamefont{Teves}},
  \bibnamefont{and}
  \bibinfo{author}{\bibfnamefont{R.}~\bibnamefont{Zukanovich~Funchal}},
  \bibinfo{journal}{Phys. Rev.} \textbf{\bibinfo{volume}{D64}},
  \bibinfo{pages}{073003} (\bibinfo{year}{2001}), \eprint{hep-ph/0105196}.

\bibitem{Gonzalez-Garcia:2001mp}
\bibinfo{author}{\bibfnamefont{M.~C.} \bibnamefont{Gonzalez-Garcia}},
  \bibinfo{author}{\bibfnamefont{Y.}~\bibnamefont{Grossman}},
  \bibinfo{author}{\bibfnamefont{A.}~\bibnamefont{Gusso}}, \bibnamefont{and}
  \bibinfo{author}{\bibfnamefont{Y.}~\bibnamefont{Nir}},
  \bibinfo{journal}{Phys. Rev.} \textbf{\bibinfo{volume}{D64}},
  \bibinfo{pages}{096006} (\bibinfo{year}{2001}), \eprint{hep-ph/0105159}.

\bibitem{Huber:2001de}
\bibinfo{author}{\bibfnamefont{P.}~\bibnamefont{Huber}},
  \bibinfo{author}{\bibfnamefont{T.}~\bibnamefont{Schwetz}}, \bibnamefont{and}
  \bibinfo{author}{\bibfnamefont{J.~W.~F.} \bibnamefont{Valle}},
  \bibinfo{journal}{Phys. Rev. Lett.} \textbf{\bibinfo{volume}{88}},
  \bibinfo{pages}{101804} (\bibinfo{year}{2002}), \eprint{hep-ph/0111224}.

\bibitem{Ota:2001pw}
\bibinfo{author}{\bibfnamefont{T.}~\bibnamefont{Ota}},
  \bibinfo{author}{\bibfnamefont{J.}~\bibnamefont{Sato}}, \bibnamefont{and}
  \bibinfo{author}{\bibfnamefont{N.-a.} \bibnamefont{Yamashita}},
  \bibinfo{journal}{Phys. Rev.} \textbf{\bibinfo{volume}{D65}},
  \bibinfo{pages}{093015} (\bibinfo{year}{2002}), \eprint{hep-ph/0112329}.

\bibitem{Huber:2002bi}
\bibinfo{author}{\bibfnamefont{P.}~\bibnamefont{Huber}},
  \bibinfo{author}{\bibfnamefont{T.}~\bibnamefont{Schwetz}}, \bibnamefont{and}
  \bibinfo{author}{\bibfnamefont{J.~W.~F.} \bibnamefont{Valle}},
  \bibinfo{journal}{Phys. Rev.} \textbf{\bibinfo{volume}{D66}},
  \bibinfo{pages}{013006} (\bibinfo{year}{2002}), \eprint{hep-ph/0202048}.

\bibitem{Ota:2004ti}
\bibinfo{author}{\bibfnamefont{T.}~\bibnamefont{Ota}} \bibnamefont{and}
  \bibinfo{author}{\bibfnamefont{J.}~\bibnamefont{Sato}}
  (\bibinfo{year}{2004}), \eprint{hep-ph/0410408}.

\bibitem{Lindner:2001fx}
\bibinfo{author}{\bibfnamefont{M.}~\bibnamefont{Lindner}},
  \bibinfo{author}{\bibfnamefont{T.}~\bibnamefont{Ohlsson}}, \bibnamefont{and}
  \bibinfo{author}{\bibfnamefont{W.}~\bibnamefont{Winter}},
  \bibinfo{journal}{Nucl. Phys.} \textbf{\bibinfo{volume}{B607}},
  \bibinfo{pages}{326} (\bibinfo{year}{2001}), \eprint{hep-ph/0103170}.

\end{thebibliography}
\end{document}